\documentclass[twocolumn,prl,preprintnumbers,amsmath,amssymb,groupedaddresst]{revtex4}
\usepackage{graphicx,ulem}

\usepackage{bm}
\newcommand{\p}{\partial}

\begin{document}

\preprint{OU-HET-866, YGHP-15-03}

\title{Speed Limit in Internal Space of Domain Walls \\
via All-Order Effective Action of Moduli Motion}

\author{Minoru Eto$^{1}$}
\author{Koji Hashimoto$^{2}$}
\affiliation{$^{1}${\it Department of Physics, Yamagata University, Kojirakawa-machi 1-4-12, Yamagata,
Yamagata 990-8560, Japan}}
\affiliation{$^{2}${\it Department of Physics, Osaka University,
Toyonaka, Osaka 560-0043, Japan}}

\begin{abstract}
We find that motion in internal moduli spaces of generic domain walls has an upper bound
for its velocity. Our finding is based on our generic formula for 
all-order effective actions of internal moduli parameter
of domain wall solitons. It is known that the 
Nambu-Goldstone mode $Z$ associated with spontaneous breaking of translation symmetry 
obeys a Nambu-Goto effective 
Lagrangian $\sqrt{1 - (\partial_0 Z)^2}$ detecting the speed of light ($|\partial_0 Z|=1$) in the target spacetime. 
Solitons can have internal moduli parameters as well, associated with a breaking of internal symmetries
such as a phase rotation acting on a field. 
We obtain, for generic domain walls, an effective Lagrangian of the internal moduli 
$\epsilon$ to all order in $(\partial \epsilon)$. The Lagrangian is given by a function of the Nambu-Goto Lagrangian: 
$L = g(\sqrt{1 + (\partial_\mu \epsilon)^2})$. This shows generically the existence of
an upper bound on $\partial_0 \epsilon$,  {\it i.e.} 
a speed limit in the internal space.  The speed limit exists even for solitons in some
non-relativistic field theories, where we find that $\epsilon$ is a type I Nambu-Goldstone mode which also obeys a nonlinear
dispersion to reach the speed limit. This offers a possibility of detecting the speed limit in condensed matter experiments. 
\end{abstract}

\maketitle


\vspace{3mm}
\noindent
{\it Introduction.} ---
Solitons are used everywhere in physics, ranging from elementary particle physics where 
domain walls are used for brane-world scenarios, to condensed matter physics where walls dividing two phases can be observed at every scale.
The dynamics of solitons is governed by low energy modes propagating on the solitons,
which are often identified as Nambu-Goldstone massless modes.
Thus providing a full effective action of Nambu-Goldstone modes on solitons is quite important
to characterize any physics in phases with broken symmetries.

It is known that a Nambu-Goto action \cite{Nambu:1986ze,Goto:1971ce}
governs the Nambu-Goldstone mode associated with a spontaneous breaking of translational symmetry.
For example, any domain wall in a flat spacetime in any dimensions supported by a 
relativistic 
field theory has its effective action of the Nambu-Goto form in a static gauge
\cite{Nielsen:1973cs,VS}
\begin{eqnarray}
L_{\rm dw} = - {\cal T}_{\rm dw} \sqrt{1 + (\partial_\mu Z)^2}\, ,
\label{NG}
\end{eqnarray}
where $\mu = 0,1,\cdots, d-1$ spans the domain wall worldvolume coordinates, 
and $Z$ is the Nambu-Goldstone mode for the translation. The value of $Z$ 
shows the location of the domain wall.
The action is valid to all order in $\partial Z$, when higher 
$(\partial)^n Z$ for  $n\geq 2$
is ignored.

The Nambu-Goto 
action reflects the special relativity of the spacetime in which the domain wall lives.
The effective action (\ref{NG}), with the wall velocity $v \equiv \partial_0 Z$ in the transverse direction, is equal to
an action of a relativistic particle, $-\sqrt{1-v^2}$. So the structure of the effective action
shows that the upper limit of the domain wall motion is the speed of light, $|v|=1$.

\begin{figure}
\includegraphics[width=7.5cm]{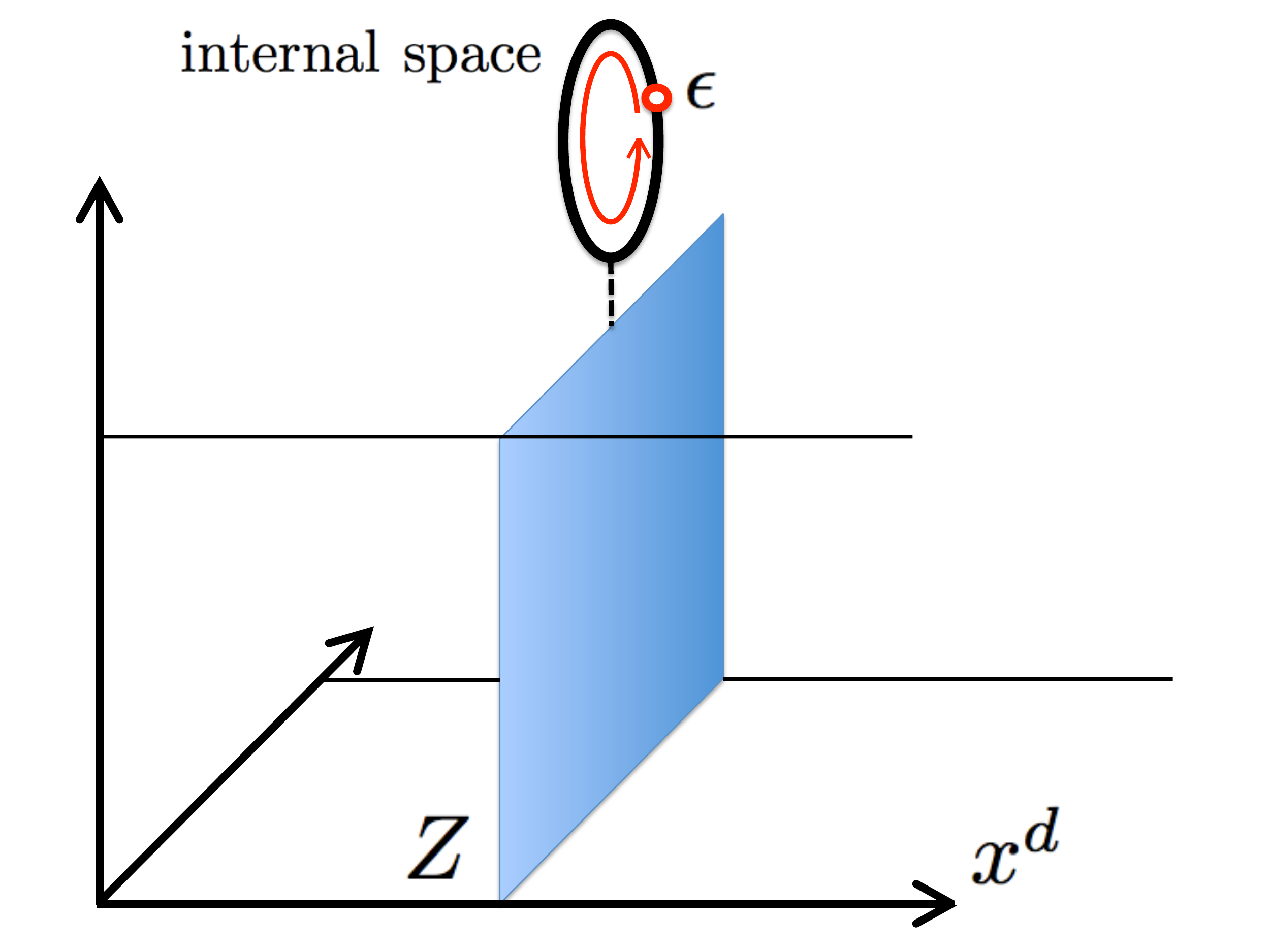}
\caption{A domain wall with an internal moduli parameter (shown as a point on
a circle above the wall). The motion of the internal moduli $\epsilon$ 
probes the internal ``space."}
\label{fig1}
\end{figure}

Internal symmetries are indispensable in physics. 
When an internal symmetry is 
broken by a domain wall, associated 
Nambu-Goldstone modes appear (See Fig.~\ref{fig1}).
The mode, called an internal moduli, is governed by some effective action.
So far, little is known for an all-order expression of the effective action, since normally 
one employs so-called Manton's method \cite{Manton:1981mp,Manton:2004tk} 
(the moduli approximation) to calculate the action order by order.

In this letter, we provide a generic form of the effective action of an internal moduli
parameter $\epsilon$ of a generic domain wall to all order in $\partial \epsilon$.
Note that the Nambu-Goto action (\ref{NG}) for $Z$ is originally given by a posteriori assumption that 
the action corresponds to the
relativistic volume of the domain wall.
On the contrary, since we could not expect such a principle to exist for the internal moduli, 
deriving the effective action might involve complicated computations dependent on details of each model.
Nevertheless, we will find that the effective action of the internal moduli 
is generically written as a 
function of the 
Nambu-Goto Lagrangian,
\begin{eqnarray}
L_{\rm dw} = g\left(m\sqrt{1 + (\partial_\mu \epsilon)^2}\right) \, ,
\end{eqnarray}
where $g$ is a system-dependent function, and $m$ is the mass of the original theory.
Thus, our result pave the way to construct the effective actions in generic models.
The action provides generically an upper limit of the moduli motion 
$\partial_0 \epsilon$. Therefore, the internal space of domain walls has
a speed limit: {\it ``an internal speed of light."}

Interestingly, our generic strategy can also be applied to non-relativistic models 
which frequently appear as effective theories  in condensed matter systems. 
Recently non-relativistic Nambu-Goldstone modes such as magnons
attracted much attention
\cite{Nielsen:1975hm,Nambu,Watanabe:2011ec,Watanabe:2012hr,Hidaka:2012ym}.
Due to our all-order effective action, the internal moduli has
a Type I dispersion for small velocity, while has a speed limit due to the 
nonlinear dispersion. The speed limit may be observed in experiments with symmetry-broken orders, such as magnetic domain walls.


\vspace{3mm}
\noindent
{\it Generic effective action in relativistic theory.} ---
We consider a generic Lagrangian of a relativistic 
complex scalar field $\phi(x^\mu,z)$ 
in $d+1$ dimensional spacetime ($\mu= 0,1,\cdots,d-1$ and $x^d = z$). 
The Lagrangian is assumed to have only two derivatives at maximum, for simplicity;
\begin{eqnarray}
{\cal L} = - F(|\phi|^2)\left(|\partial_\mu \phi|^2  +|\partial_z \phi|^2
+ m^2 |\phi|^2\right) - V(|\phi|^2) \, ,
\label{S4}
\end{eqnarray}
with generic $F$ and $V$. The field $\phi$ has a mass $m$.
This system has a $U(1)$ global symmetry which rotates the phase of the field,
$\phi \rightarrow e^{i m \epsilon_0} \phi$ where $\epsilon_0$ 
is a constant real parameter for the internal space $S^1$.
We assume the existence of a static domain wall solution with an $S^1$ moduli parameter \footnote{
This, in particular, means that the vacua, $\phi(z=-\infty)$ and $\phi(z = \infty)$,
have to be a fixed point of the $U(1)$ symmetry. Otherwise the $U(1)$ rotation
changes the vacuum and so the moduli becomes non-normalizable, which means
there is no sense in discussing effective action of the moduli parameters.}
which solves the equation of motion of (\ref{S4}):
\begin{eqnarray}
\phi = e^{i m\epsilon_0}\phi_0(z;m) \, .
\label{epsol}
\end{eqnarray}
A motion of the internal moduli parameter $\epsilon$ can be encoded as
$\epsilon = \epsilon_\mu x^\mu$ with a constant vector $\epsilon_\mu$,
which amounts to ignoring $\partial\partial \epsilon$. With this spacetime-dependent 
internal moduli parameter $\epsilon$, the exact solution is 
\begin{eqnarray}
\phi = e^{i m\epsilon}\,\phi_0\left(z;m\sqrt{1 + (\epsilon_\mu)^2}\right) \, .
\label{epsol2}
\end{eqnarray}
Notice that in the solution the mass dependence is replaced by that with
a peculiar factor $\sqrt{1 + (\epsilon_\mu)^2}$ so that the equation of motion is
satisfied. Substituting the solution (\ref{epsol2}) to the action (\ref{S4}), we obtain
the effective action of the internal moduli $\epsilon(x)$ of a generic domain wall,
\begin{eqnarray}
S_{\rm dw} &\equiv& \int d^dx dz \, 
{\cal L}\biggm|_{\phi = e^{i m\epsilon}\,\phi_0\left(z;m\sqrt{1 + (\epsilon_\mu)^2}\right) }
\nonumber 
\\
&=& \int d^dx \; g\left(m\sqrt{1 + (\partial_\mu \epsilon)^2} \right),
\label{Sdw}
\end{eqnarray}
where $g(m)$ is the on-shell action of the original static domain wall, 
$g(m)\equiv  \int dz\, {\cal L}|_{\phi = \phi_0(z;m)}$.
Interestingly, the effective action 
(\ref{Sdw}) is  {\it not} a Nambu-Goto type $\sqrt{1 + (\partial_\mu \epsilon)^2}$, 
but generically
{\it a function of the Nambu-Goto}.


\vspace{3mm}
\noindent
{\it Generic effective action in non-relativistic theory.} ---
Condensed matter systems can be approximated by a complex scalar field as an order parameter, while its theory is non-relativistic. 
Suppose the following form \footnote{The factor in the 
time component differs from the non-relativistic action given in \cite{Kobayashi:2014xua}.} of the scalar field Lagrangian,
\begin{eqnarray}
{\cal L}^{\text{(NR)}} = - F(|\phi|^2)\left(
\frac{i m_0}{2}\left(\bar{\phi}\partial_0 \phi - \phi \partial_0 \bar{\phi}\right)
+ |\partial_i \phi|^2  
\right. \nonumber \\
+|\partial_z \phi|^2
+ m^2 |\phi|^2\biggm) - V(|\phi|^2) \, ,
\label{S4nonrela}
\end{eqnarray}
with  $i=1,2,\cdots, d-1$,
which can be obtained 
from the relativistic Lagrangian (\ref{S4}) by inclusion of 
a chemical potential with a certain scaling.
The static domain wall (\ref{epsol}) remains as a solution, 
while an exact solution with moving moduli 
is given by 
\begin{eqnarray}
\phi = e^{i m\epsilon}\,\phi_0\left(z;m\sqrt{1 - \frac{m_0}{m}\p_0\epsilon + (\p_i\epsilon)^2}\right) \, .
\label{epsol3}
\end{eqnarray}
Then we can repeat the same argument to arrive at the domain wall effective action
\begin{eqnarray}
S_{\rm dw}^{\text{(NR)}} = \!\int \!d^dx \; g\!
\left(m \sqrt{1 \!-\! \frac{m_0}{m}\partial_0 \epsilon \!+\! (\partial_i \epsilon)^2}\right) \, .
\label{non-relS}
\end{eqnarray}
The effective Lagrangian is invariant under the charge conjugation $\epsilon \rightarrow
-\epsilon$ with the time reversal, as in the original Lagrangian (\ref{S4nonrela}).


\vspace{3mm}
\noindent
{\it Speed limit in internal space.} ---
The Hamiltonian for the relativistic case is calculated as
\begin{eqnarray}
H = 
-\frac{m (\partial_0 \epsilon)^2}{\sqrt{1 + (\partial_\mu\epsilon)^2}}
g' - g
\end{eqnarray}
which generically diverges at
\begin{eqnarray}
(\partial_0 \epsilon)^2 - (\partial_i \epsilon)^2 = 1 \, .
\label{sp}
\end{eqnarray}
Reaching the speed limit (\ref{sp})
expends infinite energy, as in the case of the speed of light in our spacetime.
Therefore, the internal motion has the speed limit (\ref{sp}).
The speed limit exists essentially owing to the higher order corrections in $\p\epsilon$.
In fact it cannot be seen in the usual moduli approximation to a finite order in $\p \epsilon$.
Notice that the normalization of the moduli parameter is given by $e^{im\epsilon}$. 
So the speed limit in the internal space is given 
by $m$, the mass of the original scalar field.

The speed limit sounds quite counter-intuitive, since 
the phase rotation acting on a field, $\phi \rightarrow 
e^{i m\epsilon} \phi$, can be arbitrarily fast in principle. 
It corresponds to the energy of a plane wave 
of $\phi$ 
which should not have any upper bound. However,
the physical reason of the existence of the speed limit is the
stability of the domain wall. 
Normally, the mass $m$ is 
related to the energy levels of the nonzero modes on the domain wall,
thus putting energy in the internal zero mode on the domain wall 
more than $m$ means exciting too much nonzero modes, leading to a demolition of the wall itself.

We notice that the effective Hamiltonian in the non-relativistic case derived from (\ref{non-relS}) also diverges
at 
\begin{eqnarray}
1 - (m_0/m)\partial_0 \epsilon + (\partial_i \epsilon)^2 = 0\, .
\label{spnonrela}
\end{eqnarray}
Therefore, there again exists a speed limit. In particular, the speed limit
is $\partial_0(m\epsilon) = m^2/m_0$ for $\partial_i \epsilon = 0$. 
Note that the speed limit exists only for a certain direction of motion,
and there is no speed limit at $\partial_0 (m\epsilon) = - m^2/m_0$. 
It is intriguing that even within a non-relativistic theory the domain wall internal space
can have a speed limit, 
which can be tested in experiments realizing the domain wall with an 
internal degree of freedom.

\begin{figure}
\includegraphics[width=8cm]{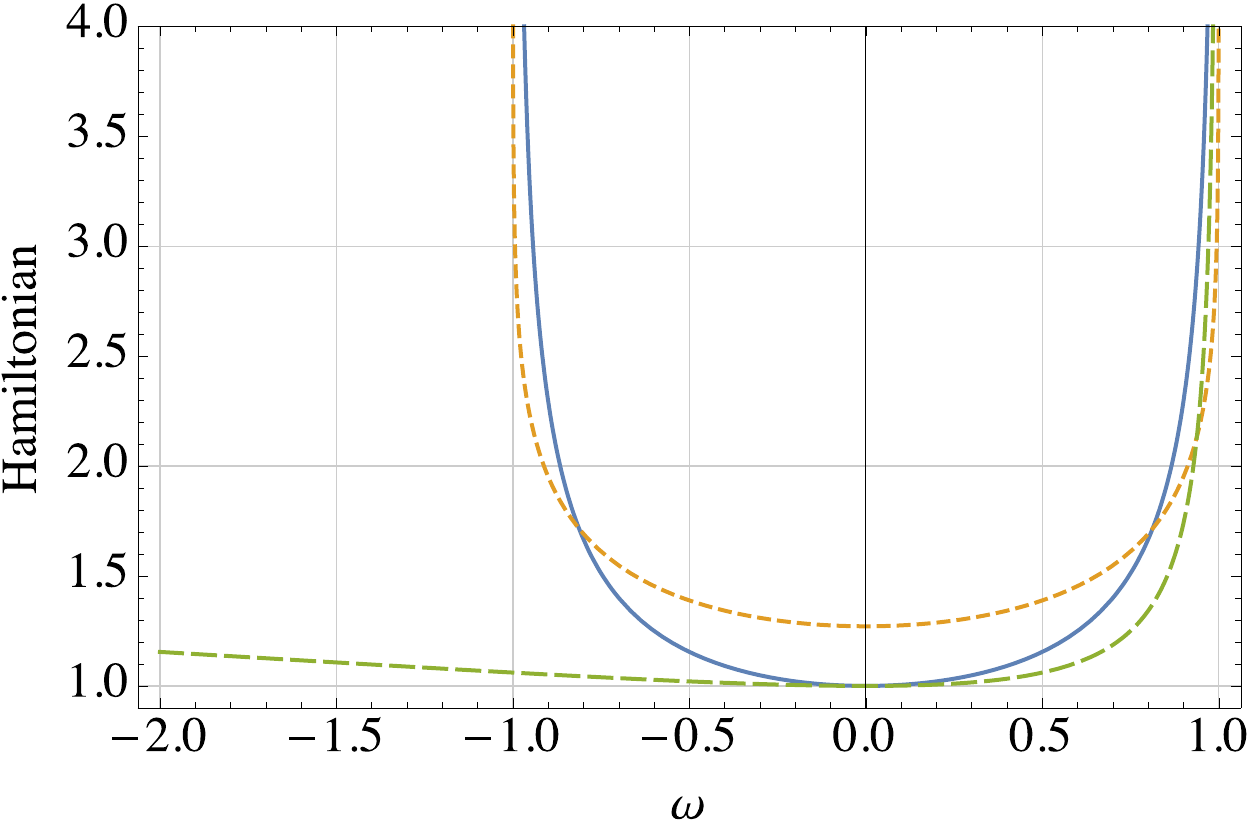}
\caption{A plot of the effective Hamiltonian of a domain wall, as a function of
$\omega \equiv \partial_0 \epsilon$ for $\partial_i \epsilon = 0$. The solid line: relativistic
${\mathbb C}P^1$ sigma model (giving a Nambu-Goto). 
The dashed line: relativistic modified ${\mathbb C}P^1$ sigma model ($\lambda = 1/2$). 
Long dashed line: non-relativistic ${\mathbb C}P^1$ sigma model ($m_0=1$). The hamiltonians ($m=1$) diverge at the speed limit (\ref{sp}) and (\ref{spnonrela}).}
\label{fig3} 
\end{figure}


\vspace{3mm}
\noindent
{\it Generalized Nambu-Goto via domain wall.} --- 
Only a  special class of Lagrangians leads to a Nambu-Goto effective action.
The condition for having a Nambu-Goto effective action in (\ref{Sdw}) is
$g(m) \propto m$. This is equivalent to having $V(|\phi|^2)=0$ in the
original Lagrangian (\ref{S4}), due to a scaling symmetry in ${\cal L}$.

A particular example which leads to a Nambu-Goto effective action
is a massive ${\mathbb C}P^1$ sigma model with the Fubini-Study metric, 
\begin{eqnarray}
F(|\phi|^2) = (1 + |\phi|^2)^{-2} \, , \quad V = 0 \, .
\label{cp1}
\end{eqnarray}
The explicit domain wall solution is $\phi_0 = e^{m z}$, and
the on-shell action is $g(m)= -m$. Thus the effective action 
of the internal moduli $\epsilon$ is given by a Nambu-Goto \cite{Eto:2015vsa},
\begin{eqnarray}
S^{\rm (massive \;  {\mathbb C}P^1)}_{\rm dw} 
= -m \int d^dx \; \sqrt{1 + (\partial_\mu \epsilon)^2} \, .
\label{CP1s}
\end{eqnarray}
Fig.\ref{fig3} is a plot of the Hamiltonian of the Nambu-Goto system (\ref{CP1s})
for $\epsilon_\mu = \omega \delta_{0\mu}$. 
It diverges at the speed limit in the internal
space, $\omega = \pm m$.

With a nontrivial $V$, we can derive various effective action of the internal space.
As an interesting example, we introduce the following $V$ in the massive 
${\mathbb C}P^1$ model,
\begin{eqnarray}
V=  4\lambda  |\phi|^2(1-|\phi|^2)^2 (1+|\phi|^2)^{-4}.
\label{V_A}
\end{eqnarray}
By the redefinition $\phi = e^{i\Phi} \tan\frac{\Theta}{2}$, a part of the massive ${\mathbb C}P^1$ model can be recast to a sine-Gordon model
with $m^2\sin^2\Theta$ potential term. The additional potential (\ref{V_A}) gives  $\lambda \sin^2 2\Theta$. Thus, the massive ${\mathbb C}P^1$
model with  (\ref{V_A}) includes as a part the so-called double sine-Gordon model which has been investigated for soliton confinement phenomena.
The static domain wall solution is
\begin{eqnarray}
\phi_0(z;m,\lambda) = 
\left[\frac
{\sqrt{4\lambda \!+\! m^2\cosh^2(\tilde{m}z)} \!+\! m^2
\sinh^2(\tilde{m}z)}
{\sqrt{4\lambda \!+\! m^2\cosh^2(\tilde{m}z)} \!-\! m^2
\sinh^2(\tilde{m}z)}\right]^{\frac{1}{2}}
\nonumber
\end{eqnarray}
with 
$\tilde{m} \equiv \sqrt{4\lambda + m^2}$, and using this, we obtain
the effective action for $\epsilon$ as
\begin{eqnarray}
S_{\rm dw} = -\int d^dx \;
\left(
\sqrt{\tilde{\lambda}} + \frac{L_{\rm NG}^2}{4\sqrt{\lambda}}
 \tanh^{-1} \sqrt{\frac{\lambda}{\tilde{\lambda}}}
\right),
\end{eqnarray}
where $\tilde{\lambda} \equiv
\lambda + L_{\rm NG}^2/4$, and $L_{\rm NG} \!=\! -m \sqrt{1+(\partial_\mu \epsilon)^2}$ 
is the Nambu-Goto Lagrangian. The Hamiltonian is given by
\begin{eqnarray}
H=
\sqrt{\tilde\lambda} +\frac{m^2\left(1 \!+\! (\partial_0 \epsilon)^2 \!+\! (\partial_i \epsilon)^2\right)}{4\sqrt{\lambda}} \tanh^{-1} \sqrt{\frac{\lambda}{\tilde{\lambda}}}
\, ,
\end{eqnarray}
which diverges at the speed limit (\ref{sp}), as in Fig.~\ref{fig3}.
A conserved charge for the symmetry $\epsilon \to \epsilon + \delta$ is 
\begin{eqnarray}
Q = \frac{m^2}{\sqrt{4\lambda}} \tanh^{-1}
\sqrt{\frac{4\lambda}{m^2\left(1 + (\p_\mu \epsilon)^2\right)+4\lambda}}\ \p_0\epsilon.
\end{eqnarray}
This also diverges at the speed limit.
With $\epsilon = \epsilon_\mu x^\mu$, $H$ and $Q$ coincide with
the tension and Noether charge calculated in the original non-linear sigma model,
which serves as a nontrivial consistency.

\vspace{3mm}

\noindent
{\it Type I Nambu-Goldstone mode.} ---
We found the generic form (\ref{non-relS}) of the effective action of $\epsilon$ to all order $\p\epsilon$.
In the non-relativistic massive ${\mathbb C}P^1$ case with (\ref{cp1}),
expanding (\ref{non-relS}) in $\p \epsilon$ up to quadratic order, 
we obtain
\begin{eqnarray}
L^{\text{(NR)}} =
m \bigg[1 - \frac{1}{2}\frac{m_0}{m}\dot\epsilon - \frac{1}{8}\left(\frac{m_0}{m}\dot\epsilon\right)^2 + \frac12 (\p_i \epsilon)^2 
\bigg].
\label{disp}
\end{eqnarray}
Note that one cannot obtain  the $\dot{\epsilon}^2$ term by the usual 
order-by-order moduli approximation:
we need the exact solution (\ref{epsol3}) to get (\ref{disp}).
From (\ref{disp}), the dispersion relation is given by
\begin{eqnarray}
\omega = \frac{2m}{m_0} |{\bf k}| \, .
\end{eqnarray}
Thus the Nambu-Goldstone mode $\epsilon$ 
is type I (that is, linear and relativistic)\footnote{
The analysis in \cite{Kobayashi:2014xua} shows that a mixing with $Z$ 
is important.
Our Lagrangian is different from that of \cite{Kobayashi:2014xua} and so there is no
$Z \dot{\epsilon}$ term. A possible mixing
$\dot{\epsilon}\dot{Z}$ does not change our result.
}.
Note that our speed limit $\omega (=\dot{\epsilon}) = m/m_0$ means
the upper bound for the value of the internal speed $\omega$.


\vspace{3mm}
\noindent
{\it Fattening and destroying the wall.} ---
As we anticipated, the motion of the internal moduli $\epsilon$ will destroy the
wall when it exceeds the speed limit. The effective Hamiltonian acquires an imaginary part, which signals the decay of the domain wall itself. The speed limit is given by 
the scalar mass $m$; turning on the moduli motion $\partial_0 \epsilon$
reduces the mass to $m \sqrt{1 - (\partial_0 \epsilon)^2}$. So the speed limit amounts
effectively to the massless limit in the original theory.

\begin{figure}[t]
\includegraphics[width=6cm]{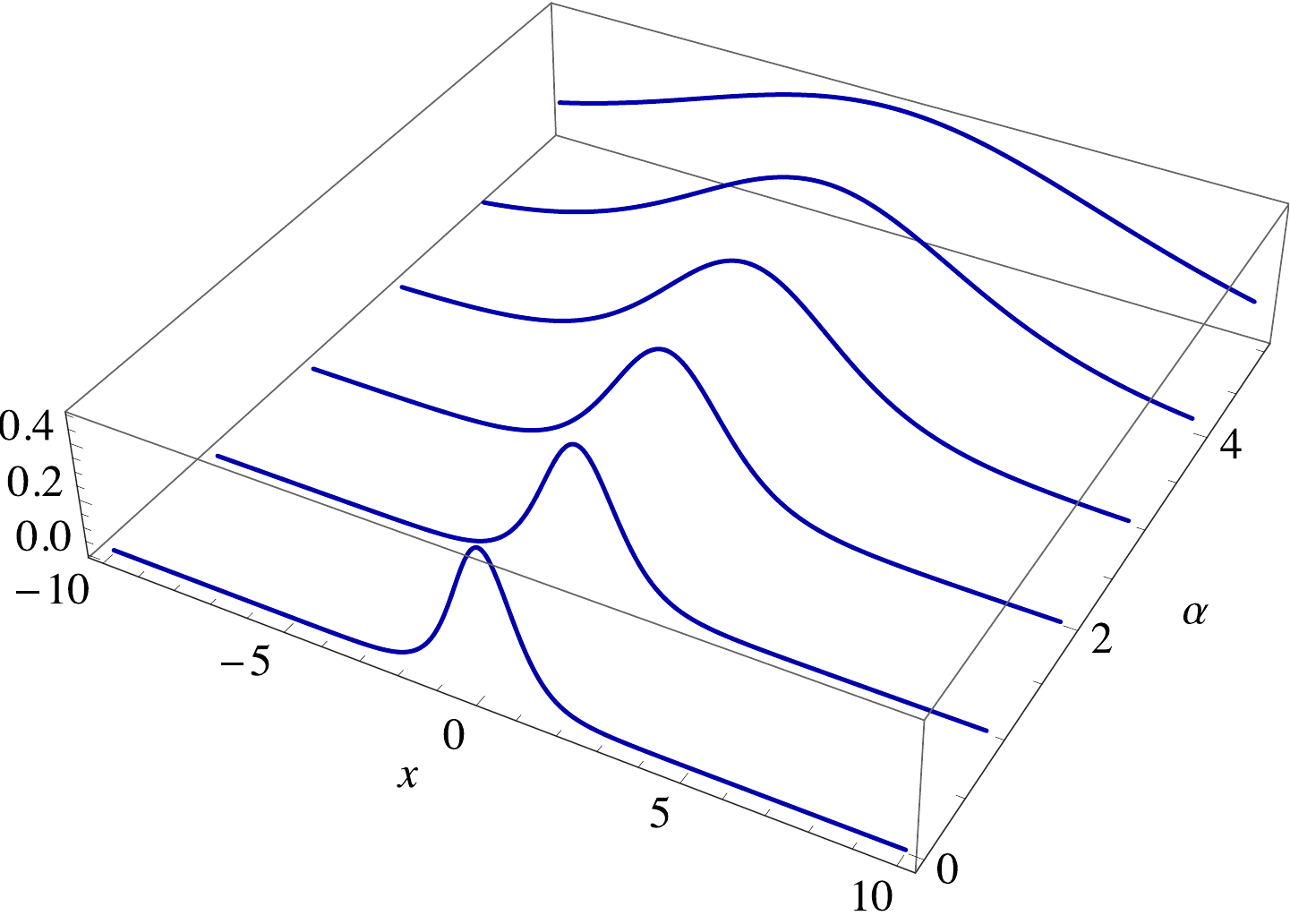}\\
\includegraphics[width=6cm]{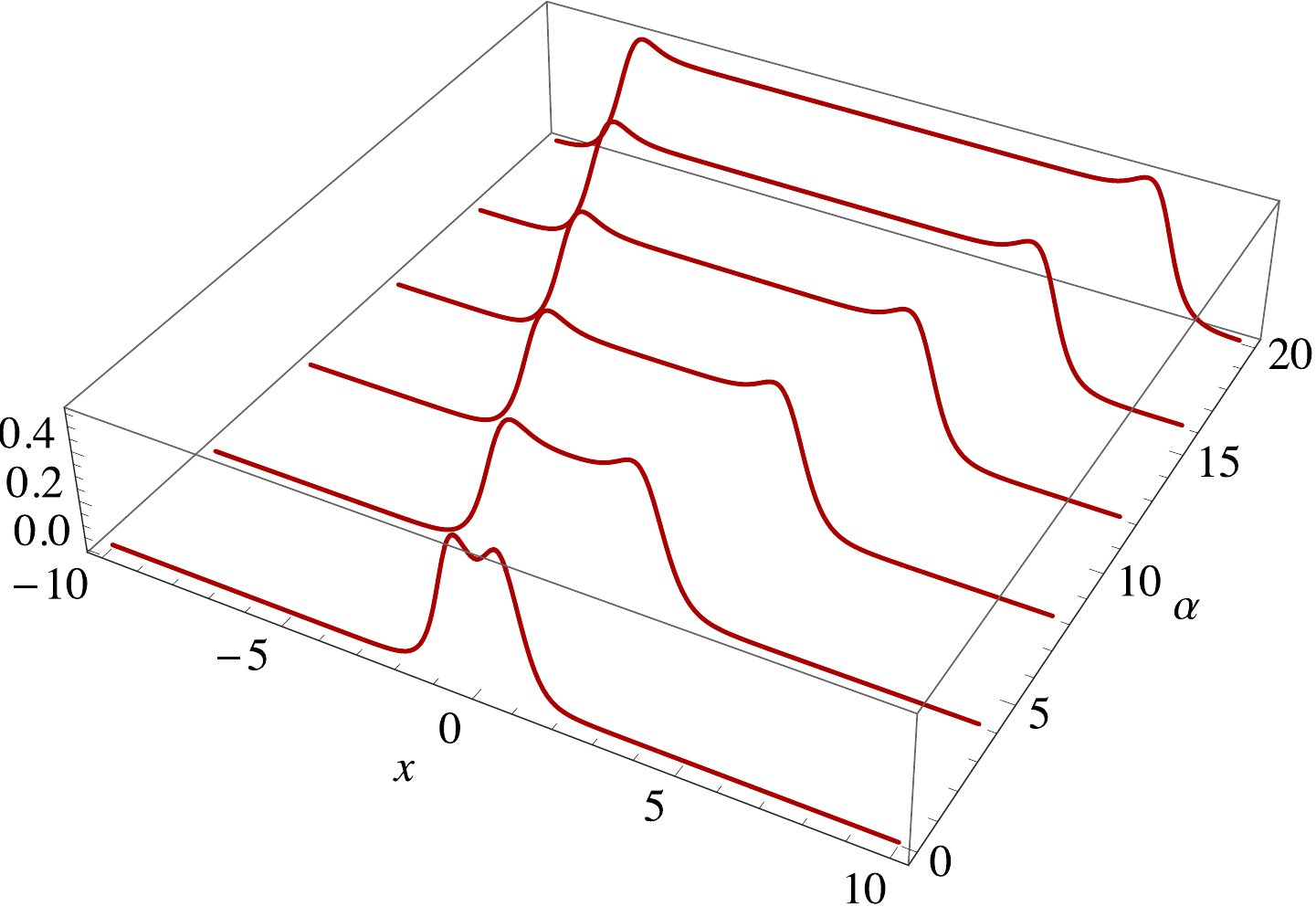}
\caption{Energy density profiles of the domain wall in
the ${\mathbb C}P^1$ with(upper)/without(lower) $V$ in (\ref{V_A}),
when we change the internal moduli $\partial_0 \epsilon$. The parameter $\alpha$
is defined by $\partial_0 \epsilon = 1- \exp[-\alpha]$. $\alpha=0$
corresponds to the original domain wall. We chose $m=1$ and $\lambda = 1/2$ for this plot.}
\label{fig4}
\end{figure}

The mass is inversely related to the width of the domain wall, so we can expect that
the internal moduli motion will make the domain wall thicker, and finally decays smoothly by the fattening. In Fig.~\ref{fig4}, 
we plot the energy density of the exact domain wall solutions in the massive ${\mathbb C}P^1$ sigma model with/without $V$ (\ref{V_A}), 
by changing $|\partial_0 \epsilon|$. 
We can clearly see the fattening of the domain walls.
In the model with $V$, at $\omega=0$ we can see that a single domain wall consists of two constituent walls connecting $\Theta = 0 \to \pi/2$ and $\pi/2 \to \pi$ respectively.
They are confined.
As $\omega$ increases, a large repulsive force appears which deconfines the constituent domain walls.

\vspace{3mm}
\noindent
{\it Relation to extra dimensions.} --- The speed limit suggests emergence of 
an extra dimension of the spacetime. In fact, we can derive the effective 
action of the internal moduli (\ref{Sdw}) by using a generalized boost along
a newly introduced extra dimension which is $S^1$. 

We upgrade the phase factor 
of the domain wall solution to an additional coordinate. 
We define a new complex scalar field $\Phi(x^M)$
which lives in the $(d+1+1)$-dimensional spacetime spanned by $x^M \equiv 
(x^\mu,z,\alpha)$,
as
\begin{eqnarray}
\Phi(x^\mu,z,\alpha) = e^{im\alpha}
\sum_{n=-\infty}^{\infty} e^{in\alpha/R} \phi^{(n)}(x^\mu,z) \, .
\end{eqnarray}
The form is written in a Fourier expansion where $n$ is the Fourier mode number
in a compact space $ 0 \leq \alpha < 2\pi R$.
We accordingly prepare for  the following 5-dimensional action
\begin{eqnarray}
S=\frac{-1}{2\pi R} \int_0^{2\pi R} \!\!\!\!\!\!\!\! d\alpha \; 
\!\!\int \!\!d^{d+1}x \, 
[F(|\phi|^2)|\partial_M \phi|^2 + V(|\phi|^2)]
\, .
\end{eqnarray}
A twisted periodicity condition (Scherk-Schwarz compactification \cite{Scherk:1979zr})
\begin{eqnarray}
\Phi(x^\mu, z, \alpha + 2\pi R) = e^{2 \pi i m R}
\Phi(x^\mu, z, \alpha) 
\label{SSb}
\end{eqnarray}
ensures the equivalence to the original system (\ref{S4}) if one ignores higher Fourier modes.
The original domain wall solution $\phi=\phi_0(z)$ is translated to
a solution in higher dimensions,
$\Phi = e^{i m \alpha} \phi_0(z)$.
Since the higher-dimensional action is Lorentz-invariant even in the extra dimension,
we can make a Lorentz boost to generate a new solution, in which the
soliton moves in the extra dimension, {\it i.e.} the internal space. Carefully
fixing an inconsistency with the boundary condition (\ref{SSb}) by a rescaling of the mass parameter, we can evaluate the on-shell action for the boosted solution and
derive the effective action (\ref{Sdw}).


\vspace{3mm}
\noindent
{\it Conclusion.} ---We show generically the existence of a speed limit in
internal moduli space $\epsilon$ of domain walls. The speed dependence of the
Hamiltonian is calculated to all order in $\partial \epsilon$.
The effective Lagrangian is generically a function of Nambu-Goto Lagrangian,
in contrast to the transverse moduli $Z$ obeying a Nambu-Goto. 
Even for non-relativistic field theory, we find the speed limit, which may be seen in experiments with symmetry-broken orders.

Our calculation can extend to other spices of solitons. In \cite{Tong:2005un}, 
D.~Tong studied 
the internal $S^1$ moduli of a 't\,Hooft Polyakov BPS monopole and showed that
it obeys a Nambu-Goto action.
Note that, as we demonstrated, the phase of the domain wall in the massive ${\mathbb C}P^1$ model also obeys 
the Nambu-Goto action. This is not a coincidence. 
Indeed, 't\,Hooft-Polyakov monopole in the Higgs phase is identified 
with a kink in the massive ${\mathbb C}P^1$ model \cite{Tong:2003pz}.
Thus, the kink-monopole correspondence may be valid to all order in $\p \epsilon$.

Furthermore, 
even in the absence of solitons, a vacuum itself can have an internal moduli,
where our method can be applied to show the non-existence of the speed limit.

When the worldvolume of the domain wall is (1+2)-dimensional, we can take a dual 
of the generalized Nambu-Goto action to obtain a general nonlinear electrodynamics.
Those generalization of the Born-Infeld action may possess interesting structure, 
including a possible relation to D-branes \cite{Abraham:1992vb,Gauntlett:2000de} in string theory and an open string metric \cite{Gibbons:2000xe,Gibbons:2002tv}.

The existence of the speed limit in internal space 
is encouraging for braneworld scenario and possible cosmological models using
a speed limit of inflation rolling \cite{Silverstein:2003hf}. 
Various applications 
to particle physics, cosmology and condensed matter physics are expected.


\vspace*{2mm}
{\noindent \it Acknowledgment.} ---
This work is dedicated to the memory of prof.~Yoichiro Nambu.
The work of M.~E.~is supported in part by JSPS KAKENHI Grant Numbers 26800119.
The work of K.~H.~is supported in part by JSPS KAKENHI Grant Numbers 15H03658, 15K13483.


\widetext
\clearpage
\begin{center}
\textbf{\large Supplementary Material}
\end{center}
\setcounter{equation}{0}
\setcounter{figure}{0}
\setcounter{table}{0}
\setcounter{page}{1}
\makeatletter
\renewcommand{\theequation}{S\arabic{equation}}
\renewcommand{\thefigure}{S\arabic{figure}}
\renewcommand{\bibnumfmt}[1]{[S#1]}

In this supplementary material, we provide detailed calculations which are used in the main part of the present article, and some additional calculations which further support the claim in the present article.
First, we show a calculation to obtain the generic effective actions
(\ref{Sdw}) and (\ref{non-relS}) in details.
Then, we provide a brief review of the derivation of the Nambu-Goto action for the translational moduli (\ref{NG}). We take a route different from that in 
\cite{Nielsen:1973cs,VS}.
Then, we re-derive our main effective action of the internal moduli (\ref{Sdw}) by using extra dimensions, as an instructive exercise. 
After that, we present detailed calculations of explicit examples of the domain wall solutions, 
as well as a new example for a composite domain wall. Two more examples follow, 
an example of a symmetry breaking at a vacuum (rather than a domain wall), and an
example of a 't\,Hooft Polyakov monopole.

\subsection{Some details for deriving the generic effective action}

In this subsection, we show details for deriving the effective action (\ref{Sdw}) and (\ref{non-relS}).
Let us begin with the equations of motion of the relativistic Lagrangian (\ref{S4}),
\begin{eqnarray}
-\p_\mu(F\p^\mu\phi) -\p_z(F\p^z\phi) + m^2 F' |\phi|^2 \phi + m^2 F \phi + V'\phi + F' \phi |\p_\mu\phi|^2  + F' \phi |\p_z\phi|^2 = 0.
\label{eom}
\end{eqnarray}
The domain wall solution perpendicular to the $z$ axis
\begin{eqnarray}
\phi(z) = e^{im\epsilon_0} \phi_0(z;m),
\label{sol_static}
\end{eqnarray}
is a solution to the reduced equation
\begin{eqnarray}
-\p_z(F\p_z\phi_0) + m^2 (F' |\phi_0|^2 + F) \phi_0 + V'\phi_0 + F' \phi_0 |\p_z\phi_0|^2= 0.
\label{eom_static}
\end{eqnarray}
Let us make an ansatz for a generic solution
\begin{eqnarray}
\phi = e^{im \epsilon}\varphi_0(z),\quad (\epsilon = \epsilon_\mu x^\mu).
\label{sol_gene}
\end{eqnarray}
Plugging this into Eq.~(\ref{eom}), we  find that $\varphi$ obeys the following equation
\begin{eqnarray}
-\p_z(F\p_z\varphi_0) + m^2(1+(\epsilon_\mu)^2) (F' |\varphi|^2 + F) \varphi_0 + V'\varphi_0 + F' \varphi_0 |\p_z\varphi_0|^2 = 0.
\end{eqnarray}
Comparing this with (\ref{eom_static}), we can regard $\varphi$ as a static solution in the model (\ref{S4}) with
$m$ being replaced with $m \sqrt{1+(\epsilon_\mu)^2}$.
Namely, we get the generic solution
\begin{eqnarray}
\varphi_0(z) = \phi_0\!\left(z;m\sqrt{1+(\epsilon_\mu)^2}\right).
\end{eqnarray}
Correspondingly, if we plug $\phi(x^\mu,z) = e^{im\epsilon}\varphi(x^\mu,z)$ 
into the Lagrangian (\ref{S4}), we find the Lagrangian for $\varphi$
\begin{eqnarray}
\tilde {\cal L} = -F(|\varphi|^2)\left(|\p_\mu\varphi|^2 + |\p_z\varphi|^2 + m \epsilon_\mu j^\mu + m^2\left(1 + (\epsilon_\mu)^2)\right)|\varphi|^2\right) - V(|\varphi|^2),
\end{eqnarray}
with $j^\mu = i (\varphi\p^\mu \varphi^* - \varphi^*\p^\mu \varphi)$.
Therefore, the on-shell Lagrangian for the generic solution (\ref{sol_gene}) is directly obtained as
\begin{eqnarray}
\int dz\ {\cal L}(\phi=e^{im\epsilon}\varphi_0) = 
\int dz\ \tilde{\cal L} (\varphi_0) = g \left(m \sqrt{1+(\epsilon_\mu)^2}\right),
\end{eqnarray}
with $g(m) = \int dz\ {\cal L}(\phi = e^{im\epsilon_0}\phi_0(z;m))$.
Finally, just replacing $\epsilon_\mu$ by $\p_\mu \epsilon$, we reach at the final result (\ref{dwres}).

We can repeat the same argument for the non-relativistic model.
The equations of motion for the non-relativistic Lagrangian (\ref{S4nonrela}) is given as
\begin{eqnarray}
\frac{im_0}{2}\p^0(F\phi)- \p_I (F\p_I\phi)  + m^2 (F' |\phi|^2 + F) \phi + V'\phi + \frac{im_0}{2}\left\{F'(\bar\phi\p_0\phi - \phi\p_0\bar\phi)\phi
+F\p_0\phi\right\}+F'|\p_I\phi|^2 \phi = 0,
\end{eqnarray}
with $x^I=x^1,x^2,\cdots,x^{d-1},z$.
Clearly, (\ref{sol_static}) is a solution of this equation. Furthermore, the equation for $\varphi_0$ with the Ansatz (\ref{sol_gene})
is obtained as
\begin{eqnarray}
- \p_z (F\p_z\varphi_0)  + (m^2-m_0\epsilon_0 + m^2(\epsilon_i)^2) (F' |\varphi_0|^2 + F) \varphi_0 + V'\varphi_0 + F' \varphi_0 |\p_z\varphi_0|^2  = 0.
\end{eqnarray}
Comparing this with (\ref{eom_static}), we find the generic solution in the non-relativistic model is given by
\begin{eqnarray}
\varphi_0(z) = \phi_0\left(z;m\sqrt{1-\frac{m_0}{m}\epsilon_0 + (\epsilon_i)^2}\right).
\end{eqnarray}
Correspondingly, if we plug $\phi(x^\mu,z) = e^{im\epsilon}\varphi(x^\mu,z)$  into the non-relativistic Lagrangian (\ref{S4nonrela}), we find the Lagrangian for $\varphi$
\begin{eqnarray}
\tilde {\cal L}^{\text{(NR)}} = - F(|\varphi|^2)\left(
\frac{i m_0}{2}\left(\bar{\varphi}\partial_0 \varphi - \varphi \partial_0 \bar{\varphi}\right)
+ |\partial_i \varphi|^2  
+|\partial_z \varphi|^2 + m \epsilon_i j^i
+ (m^2 - m_0\epsilon_0 + (\epsilon_i)^2) |\varphi|^2\right) - V(|\varphi|^2) \, .
\end{eqnarray}
Therefore, the on-shell value of the original non-relativistic Lagrangian for the generic solution is given by
\begin{eqnarray}
\int dz\ {\cal L}^{\rm (NR)} (\phi=e^{im\epsilon}\varphi_0) = 
\int dz\ \tilde{\cal L}^{\rm (NR)}  (\varphi_0) = g \left(m\sqrt{1-\frac{m_0}{m}\epsilon_0 + (\epsilon_i)^2}\right).
\end{eqnarray}

\section{Deriving Nambu-Goto effective action for translational moduli}

Let us start with a real scalar field $\phi$ in $d+1$ dimensions. The Lagrangian 
is given by ${\cal L}$ and thus the action is written as
\begin{eqnarray}
S = \int d^{d+1}x \; {\cal L} [\phi]\, .
\end{eqnarray}
Suppose we obtain a domain wall as a classical solution of this system, 
\begin{eqnarray}
\phi = \phi_0(z),\quad (z = x^d) \, .
\label{clsol}
\end{eqnarray}
The domain wall worldvolume is perpendicular to the direction $z$, 
so it is flat along the remaining directions $x^\mu$ ($\mu=0,1,\cdots,d-1$).
The obvious zero mode $Z$ of the domain wall is the position of the domain wall
in the $z$ axis. Inclusion of the zero mode gives us a generic solution
\begin{eqnarray}
\phi = \phi_0(z-Z)
\end{eqnarray}
where $Z$ is a constant parameter. Turning on $Z$ does not cost any energy
and this  remains as a classical solution of the original system ${\cal L}$. 

Now, we are interested in the low energy effective description of the domain wall.
The zero mode $Z$ can depend on the worldvolume coordinates $x^\mu$,
as $Z(x^\mu)$.
Ignoring the higher derivatives such as $(\partial)^2 Z$, we should be able to
obtain an effective action of the domain wall
\begin{eqnarray}
S_{\rm dw} = \int dx^{d-1} \; L_{\rm dw}[\partial_\mu Z] \, .
\end{eqnarray}
The domain wall Lagrangian $L_{\rm dw}$ is a functional of $\partial_\mu Z$ only,
as the potential term $V(Z)$ should not appear because $Z$ is a zero mode.

The easiest way to get $L_{\rm dw}$ is to make a Lorentz transformation.
Consider the following Lorentz transformation
\begin{eqnarray}
\tilde{z} &=& \Lambda^d_{\;\; d} z + \Lambda^d_{\;\; \nu} x^\nu 
\label{L3} \\
\tilde{x}^\mu &=& \Lambda^\mu_{\;\; d} z + \Lambda^\mu_{\;\; \nu} x^\nu 
\label{Lm} 
\end{eqnarray}
where $\Lambda$ is an $SO(1,d)$ transformation matrix whose determinant is $1$.
Then obviously we can generate a new classical solution
\begin{eqnarray}
\phi = \phi_0(\tilde{z}).
\label{Ltr}
\end{eqnarray}
This solution is a tilted 
domain wall moving in a transverse direction with a constant velocity.
Since
\begin{eqnarray}
\tilde{z} = 
\Lambda^d_{\;\; d} \left(z + (\Lambda^d_{\;\; d})^{-1}\Lambda^d_{\;\; \mu} x^\mu\right) , 
\end{eqnarray}
it is possible to regard
\begin{eqnarray}
Z(x^\mu) = - (\Lambda^d_{\;\; d})^{-1}\Lambda^d_{\;\; \mu} x^\mu \, .
\label{Zdef}
\end{eqnarray}
So, once we obtain a domain wall effective action as a function of the Lorentz transformation
matrix elements $\Lambda$, using (\ref{Zdef})
we can regard it as a domain wall effective action for $Z(x^\mu)$. 

Let us calculate the effective action of the domain wall.
Substituting the transformed solution (\ref{Ltr}) into the original action, we can obtain
the effective action
\begin{eqnarray}
S_{\rm dw} &=&  \int d^dx \int dz \; {\cal L} \biggm|_{\phi = \phi_0(\tilde{z})} \, .
\label{Sdwf}
\end{eqnarray}
To compute this integral, we use the following trick. The new 
domain wall solution (\ref{Ltr}) depends only on $\tilde{z}$ while the
spacetime coordinates $x^\mu$ should be kept as it is, so we consider
a general coordinate transformation (note that it is not a Lorentz transformation)
\begin{eqnarray}
\tilde{z} &=& \Lambda^d_{\;\; d} z + \Lambda^d_{\;\; \nu} x^\nu \, ,
\label{L3n}
\\
\tilde{x}^\mu &=& x^\mu \, .
\label{Lmn}
\end{eqnarray}
The transformation for $z$ is the same as the Lorentz transformation (\ref{L3})
while $x^\mu$ is kept intact. For this general coordinate transformation, the Jacobian 
is found as
\begin{eqnarray}
\frac{d\tilde{x}}{dx} = \left(
\begin{array}{cc}
\delta^\mu{}_{\nu} & 0 \\
\Lambda^d{}_\nu& \Lambda^d{}_d 
\end{array}
\right) \, ,
\end{eqnarray}
so the metric
is transformed as
\begin{eqnarray}
g^{\tilde{d}\tilde{d}} = (\Lambda^d{}_d)^2 + \Lambda^d{}_\mu \Lambda^d{}_\nu \eta^{\mu\nu} \, =1 \, .
\end{eqnarray}
The last equality is due to the $SO(1,d)$ nature of the
Lorentz transformation $\Lambda$. 
Note that there appears 
nontrivial off-diagonal elements of the metric such as
$g^{\tilde{d}\mu}$, although they are irrelevant in the following calculations.

When we substitute the new solution (\ref{Ltr}) into the Lagrangian,
we may use the new Lagrangian transformed by the general coordinate transformation
(\ref{L3n}) (\ref{Lmn}), since the Lagrangian is a scalar quantity under the 
general coordinate transformation. The solution depends only on $\tilde{z}$,
so all derivatives $\p_\mu$ acting on the solution vanishes. The only terms which
are relevant are 
$g^{\tilde{d}\tilde{d}}\partial_{\tilde{d}} \phi \partial_{\tilde{d}}\phi^*$ and
its higher derivative analogues. 
Since $g^{\tilde{d}\tilde{d}}=1$, this concludes that the Lagrangian with
the new solution is equal to the Lagrangian with the original solution, via a simple replacement $z$ by $\tilde{z}$.
Then the domain wall effective action (\ref{Sdwf}) can be evaluated as
\begin{eqnarray}
\int d^dx \int dz \; {\cal L} \biggm|_{\phi = \phi_0(\tilde{z})}
=\int d^dx \int d\tilde{z} \sqrt{-\det \tilde{g}}
\left[ {\cal L} \biggm|_{\phi = \phi_0(z)}\right]_{z \text{ replaced  by } \tilde{z}}
\;.
\end{eqnarray}
The general coordinate transformation (\ref{L3n}) (\ref{Lmn}) gives
\begin{eqnarray}
\det \tilde{g} = \frac{1}{(\Lambda^d_{\;\; d})^2}\, ,
\end{eqnarray}
so we obtain 
\begin{eqnarray}
S_{\rm dw} &=&
\frac{1}{\Lambda^d_{\;\; d}}\int d^dx \int d\tilde{z}
\left[ {\cal L} \biggm|_{\phi = \phi_0(z)}\right]_{z \text{ replaced by } \tilde{z}}
\nonumber \\
&=&
\frac{1}{\Lambda^d_{\;\; d}}\int d^dx \int dz
{\cal L} \biggm|_{\phi = \phi_0(z)} \, .
\end{eqnarray}
We define
the on-shell value of the Lagrangian with the classical solution (\ref{clsol})
integrated over $z$ as a tension ${\cal T}$ of the domain wall,
\begin{eqnarray}
 {\cal T}_{\rm dw} \equiv -\int dz
\; {\cal L} \biggm|_{\phi = \phi_0(z)} \, .
\end{eqnarray}
Then we find the domain wall effective Lagrangian
\begin{eqnarray}
L_{\rm dw} = -{\cal T}_{\rm dw} \frac{1}{\Lambda^d_{\;\; d}} \, .
\end{eqnarray}
Using the $SO(1,d)$ relation and the relation (\ref{Zdef}), we find
\begin{eqnarray}
\frac{1}{(\Lambda^d_{\;\; d})^2}  = 1+   (\Lambda^d_{\;\; d})^{-2}\Lambda^d_{\;\; \mu}
\Lambda^d_{\;\; \nu} \eta^{\mu\nu}
= 1 + (\partial_\mu Z)^2 \, ,
\end{eqnarray}
so we finally obtain the domain wall effective Lagrangian
\begin{eqnarray}
L_{\rm dw} = -{\cal T}_{\rm dw} \sqrt{1 + (\partial_\mu Z)^2} \, .
\label{NGeff}
\end{eqnarray}

This is nothing but a Nambu-Goto Lagrangian. Hence we conclude that the
effective action of the translational zero mode $Z(x^\mu)$ 
of any domain wall is a Nambu-Goto 
action, up to the first derivative of the zero mode $Z(x^\mu)$.

\section{Extra dimensions and Deriving effective action for internal moduli}

Previously we derived the effective action of the internal moduli of the 
domain wall, using the equations of motion and explicit solutions.
Here, we shall utilize an embedding to a spacetime with an extra dimension,
to derive the same expressions for the effective action of the internal moduli.
%

First, we study what form of the Lagrangian can give a domain wall with 
an internal moduli space $S^1$. Next, we embed the scalar system into
a higher dimensional spacetime where the internal phase rotation is related
to the extra dimensional coordinate. We
will obtain generic
effective action of the domain wall. Then, we study what condition should be met
for the action to be Nambu-Goto type. Finally, we study the speed limit in the
internal space.

\subsection{System with a domain wall with $S^1$ internal moduli}

We consider a generic Lagrangian of a complex scalar field $\phi(x^\mu,z)$
in a flat $(d+1)$-dimensional spacetime ($\mu=0,\cdots,d-1$). 
The Lagrangian is assumed to have only two derivatives at maximum, for simplicity;
\begin{eqnarray}
S= \int d^{d+1}x \; {\cal L} \, , \quad
{\cal L} = - F(|\phi|^2)\left(|\partial_\mu \phi|^2  +|\partial_z \phi|^2
+ m^2 |\phi|^2\right) - V(|\phi|^2) \, .
\end{eqnarray}
The mass term is intendedly separated from the potential functional $V(|\phi|^2)$.
This system has a $U(1)$ global symmetry which rotates the phase of the field,
\begin{eqnarray}
\phi \rightarrow e^{i m \epsilon} \phi, \quad \epsilon \in {\mathbb R} \, .
\end{eqnarray}
We assume the existence of a domain wall solution
\begin{eqnarray}
\phi = \phi_0 (z)
\end{eqnarray}
where $\phi_0$ is a real function which interpolates two vacua which have 
the same energy. We need an $S^1$ moduli for the domain wall, so the domain wall
needs to be a solution even if we rotate it as
\begin{eqnarray}
\phi = e^{i m\epsilon}\phi_0(z) \, .
\label{epsol_S}
\end{eqnarray}
This, in particular, means that the vacua, $\phi(z=-\infty)$ and $\phi(z = \infty)$,
have to be a fixed point of the $U(1)$ symmetry. Otherwise the $U(1)$ rotation
changes the vacuum and so the moduli becomes non-normalizable, which means
there is no sense in discussing effective action of the moduli parameters.

Generically the fixed points of the phase rotation are 
only $\phi=0$ and $\phi=\infty$ ($1/\phi = 0$), so this condition of having
the $S^1$ moduli constrains the Lagrangian as follows: 
the total potential
\begin{eqnarray}
V_{\rm tot} \equiv m^2 |\phi|^2 F(|\phi|^2)  +  V(|\phi|^2)
\label{Vcond}
\end{eqnarray}
has two minima at $\phi=0$ and $\phi=\infty$. In the following, we treat this 
system. Several examples will be given in later.

\subsection{Generalized Lorentz boost creating new solutions}

Whether the internal moduli parameter can be regarded as another spatial coordinate or not 
is an important question. 
As we have seen in the previous section, the Nambu-Goto action can be obtained
by the Lorentz transformation of the soliton solution. So, here, we consider a Lorentz transformation including the internal moduli direction. We will be careful about 
what situation we can perform the Lorentz transformation, below.

First, we need to introduce the internal coordinate. Since the $S^1$ moduli is
associated with the phase $U(1)$ symmetry acting on the field $\phi$, 
it is natural to upgrade the phase factor 
of the domain wall solution to the additional spatial coordinate $\alpha$. The new internal space needs to
be compact, so we define a new complex scalar field $\Phi(x^\mu,z,\alpha)$
which lives in the $(1+d+1)$-dimensional spacetime spanned by $x^M \equiv 
(x^\mu,z,\alpha)$,
as
\begin{eqnarray}
\Phi(x^\mu,z,\alpha) = e^{im\alpha}
\sum_{n=-\infty}^{\infty} e^{in\alpha/R} \phi^{(n)}(x^\mu,z) \, .
\label{SSdef}
\end{eqnarray}
The form is written in a Fourier expansion where $n$ is the Fourier mode number.
In addition, we have introduced an overall factor $e^{im\alpha}$ by the following reason.
Let us prepare the following $(d+2)$-dimensional action
\begin{eqnarray}
S &=& \frac{1}{2\pi R} \int_0^{2\pi R} \!\!\! d\alpha \; 
\int d^dx \, dz \, {\cal L}_{d+2} \, ,
\label{S5d+2}
\\
{\cal L}_{d+2} &=& - F(|\phi|^2)|\partial_M \phi|^2 - V(|\phi|^2) \, .
\end{eqnarray}
Then, substituting the expansion (\ref{SSdef}), the factor $e^{im\alpha}$ provides
exactly the mass term in the $(d+1)$-dimensional action (\ref{S4}). In fact, when
we have only the $n=0$ mode in (\ref{SSdef}), then substituting it to (\ref{S5d+2})
exactly reproduces (\ref{S4}). The factor $e^{i m \alpha}$ in (\ref{SSdef}) manifests
a twisted periodicity condition
\begin{eqnarray}
\Phi(x^\mu, z, \alpha + 2\pi R) = e^{2 \pi i m R }
\Phi(x^\mu, z, \alpha) 
\end{eqnarray}
which is called Scherk-Schwarz compactification \cite{Scherk:1979zr}.

Now, we have the domain wall solution $\phi=\phi_0(z)$ of the original ${\cal L}$,
then
\begin{eqnarray}
\Phi = \Phi_0(\alpha,z) \equiv e^{i m \alpha} \phi_0(z)
\label{up4}
\end{eqnarray}
is a solution of the equation of motion of ${\cal L}_{d+2}$. This solution just have the $n=0$ 
component of the expansion (\ref{SSdef}). 

Since it is a solution of ${\cal L}_{d+2}$, 
and since ${\cal L}_{d+2}$ appears to be Lorentz invariant in $(d+2)$ dimensions, we can make use
of the $(d+2)$-dimensional Lorentz transformation to create a new classical solution.
In particular we are interested in the internal coordinate $\alpha$, let us make a 
transformation in the sub-spacetime spanned by $\alpha$ and $x^\mu$.
As in the previous section, the Lorentz transformation is
\begin{eqnarray}
\tilde{\alpha} &=& 
\Lambda^\alpha_{\;\; \alpha} \alpha + \Lambda^\alpha_{\;\; \mu} x^\mu \, ,
\\
\tilde{x}^\mu &=& 
\Lambda^\mu_{\;\; \alpha} \alpha + \Lambda^\mu_{\;\; \nu} x^\nu \, .
\end{eqnarray}
We act this transformation to obtain a new solution,
\begin{eqnarray}
\Phi & = & \Phi_0(\tilde{\alpha},z)
\nonumber \\
 &=& e^{i m \tilde{\alpha}} \phi_0(z)
\nonumber \\
& = & e^{i m \Lambda^\alpha_{\;\; \alpha} \alpha}
e^{i m \Lambda^\alpha_{\;\; \mu} x^\mu} \phi_0(z) \, .
\label{newsol}
\end{eqnarray}
Note that this is a solution of ${\cal L}_{d+2}$ while it is not a solution of 
the original $(d+1)$-dimensional ${\cal L}$. The reason is that to make the reduction to the $d+1$ dimensions
we need the relation (\ref{up4}) where the
dependence on the extra dimension is $e^{i m \alpha}$, 
while the new solution (\ref{newsol}) has a different phase factor 
$e^{i m \Lambda^\alpha_{\;\; \alpha} \alpha}$. 
In fact, a new $(d+1)$-dimensional solution which can be read off from (\ref{newsol}) as
\begin{eqnarray}
\phi = e^{i m \Lambda^\alpha_{\;\; \mu} x^\mu} \phi_0(z) \, 
\label{newsolphi}
\end{eqnarray}
is not a solution of ${\cal L}$ but a solution of
\begin{eqnarray}
\tilde{\cal L}= {\cal L}\biggm|_{m \rightarrow m\Lambda^\alpha_{\;\; \alpha}} \, .
\label{rep}
\end{eqnarray}
The reason is obvious. The $\alpha$ dependence in the new solution (\ref{newsol})
provides, together with the $\alpha$ derivatives in ${\cal L}_{d+2}$, a new mass term
which is $m^2 (\Lambda^\alpha_{\;\; \alpha})^2$ instead of the original $m^2$.
In other words, we define a new solution
\begin{eqnarray}
\phi = e^{i (m/\Lambda^\alpha_{\;\; \alpha})\Lambda^\alpha_{\;\; \mu} x^\mu} \phi_0(z;m/\Lambda^\alpha_{\;\; \alpha}) \, ,
\label{newp}
\end{eqnarray}
where we have replaced $m$ by $m/\Lambda^\alpha_{\;\; \alpha}$, then this 
(\ref{newp}) is a solution of the original Lagrangian ${\cal L}$. In this way, 
we can create new solutions by a generalized Lorentz transformation in the space including the internal direction.
Taking $\Lambda^\alpha_{\;\; \alpha} = 1/\sqrt{1+(\epsilon_\mu)^2}$ and
$\Lambda^\alpha_{\;\; \mu} = \epsilon_\mu/\sqrt{1+(\epsilon_\nu)^2}$, we get the solution (\ref{epsol2}).

\subsection{General effective action for the internal $S^1$ moduli}

We are ready for evaluating the effective action. Let us first calculate it 
as an effective action of a $(d+2)$-dimensional solution. (That is, we here calculate first
the effective action without the
above replacement of $m$. Later we shall incorporate the effect of the replacement.)
The on-shell action is
\begin{eqnarray}
S &=& \frac{1}{2\pi R} \int_0^{2\pi R} \!\!\! d\alpha \; 
\int d^dx \, dz \, {\cal L}_{d+2} \biggm|_{\Phi = \Phi_0(\tilde{\alpha},z)} ,
\end{eqnarray}
and to evaluate this explicitly we perform a general coordinate transformation
\begin{eqnarray}
\tilde{\alpha} = 
\Lambda^\alpha_{\;\; \alpha} \alpha + \Lambda^\alpha_{\;\; \mu} x^\mu \, ,
\quad 
\tilde{x}^\mu = 
 x^\mu \, 
\end{eqnarray}
as in the previous section. Then, using $g^{\tilde{\alpha}\tilde{\alpha}}=1$
and $\sqrt{-\det \tilde{g}} = 1/\Lambda^\alpha{}_\alpha$, we find
\begin{eqnarray}
S &=& 
\frac{1}{2\pi R} 
\int_{\Lambda^\alpha_{\;\; \mu} x^\mu}^{2\pi R\Lambda^\alpha_{\;\; \alpha} +\Lambda^\alpha_{\;\; \mu} x^\mu} \frac{d\tilde{\alpha}}{\Lambda^\alpha_{\;\; \alpha}} \; 
\int d^dx \, dz \, 
\left[{\cal L}_{d+2} \biggm|_{\Phi = \Phi_0(\alpha,z)} \right]_{\alpha\text{ replaced by } \tilde{\alpha}}
\nonumber
\\
& = &
\frac{1}{2\pi R\Lambda^\alpha_{\;\; \alpha}} 2\pi R \Lambda^\alpha_{\;\; \alpha}
\int d^dx \, dz \, {\cal L}_{d+2} \biggm|_{\Phi = \Phi_0(\alpha,z)}
\nonumber
\\
& = &
\int d^dx \, dz \, {\cal L}_{d+2} \biggm|_{\Phi = \Phi_0(\alpha,z)} \, 
\nonumber
\\
& = &
\int d^dx \,  g(m)\, .
\label{Sinv}
\end{eqnarray}
In the last equality 
we have defined the on-shell action for the original domain wall with the $(d+1)$-dimensional Lagrangian,
\begin{eqnarray}
g(m) \equiv \int dz \, {\cal L} \biggm|_{\phi = \phi_0(z)} \, .
\end{eqnarray}
(\ref{Sinv}) means that the on-shell action is independent of the Lorentz
transformation parameters appearing in the solution. 

However, we have to remember the fact that the new solution (\ref{newsolphi})
is not a solution of ${\cal L}$ but a solution of $\tilde{\cal L}$ defined in (\ref{rep})
which is
given by the replacement $m \rightarrow m\Lambda^\alpha_{\;\; \alpha}$.
So, as mentioned in (\ref{newp}), to have a solution of the original $(d+1)$-dimensional
Lagrangian, we need to make a redefinition of $m$ as $m \rightarrow
m/\Lambda^\alpha_{\;\; \alpha}$. Therefore, the correct effective action of
the domain wall described by the solution (\ref{newp}) is given by 
(\ref{Sinv}) with the replacement, 
\begin{eqnarray}
S_{\rm dw} =\int d^dx \, g(m/\Lambda^\alpha_{\;\; \alpha}) \, . 
\label{dwac}
\end{eqnarray}

We shall reinterpret the factor $1/\Lambda^\alpha_{\;\; \alpha}$ as a function
of the moduli. The procedure is already given in the previous section.
Comparing the constant moduli parameter $\epsilon$ in the solution (\ref{epsol_S})
and the new solution (\ref{newp}), we can regard the dynamical moduli parameter
to have a configuration in the new solution as
\begin{eqnarray}
\epsilon(x^\mu) = 
\Lambda^\alpha_{\;\; \mu} x^\mu/\Lambda^\alpha_{\;\; \alpha} \, .
\end{eqnarray}
This is reminiscent of (\ref{Zdef}) for the translational zero mode $Z$ in the previous section. So, similarly, we have
\begin{eqnarray}
\frac{1}{(\Lambda^\alpha_{\;\; \alpha})^2}
= 1+ (\Lambda^\alpha_{\;\; \alpha} )^{-2} \Lambda^\alpha_{\;\; \mu}
\Lambda^\alpha_{\;\; \nu} \eta^{\mu\nu}
= 1  + (\partial_\mu \epsilon)^2\, .
\end{eqnarray}
Substituting this expression to the domain wall effective action (\ref{dwac}), we obtain
the final form of the generic effective action for the internal moduli $\epsilon(x^\mu)$
of the domain wall as
\begin{eqnarray}
S_{\rm dw} = \int d^dx \, \, g\left(m\sqrt{1 + (\partial_\mu \epsilon)^2}\right) \, .
\label{dwres}
\end{eqnarray}
This precisely reproduces (\ref{Sdw}).

The generic action is a function of the Nambu-Goto action. 
In the next subsection, we learn a condition of having the Nambu-Goto form.

\subsection{The condition for having the Nambu-Goto}

Our final expression for the effective action of the internal moduli is (\ref{dwres}).
Obviously, the condition that this action becomes Nambu-Goto action is
to have a linear $g(m)=A m$ where $A$ is a constant.

We will find below that a sufficient condition to have a Nambu-Goto action
for the effective action, in other words, to have a linear $g(m)$, is 
to start with 
\begin{eqnarray}
V=0
\end{eqnarray}
in the original $(d+1)$-dimensional Lagrangian.
When $V=0$, the total action is
\begin{eqnarray}
S= - \int d^{d+1}x \;  F(|\phi|^2)\left(|\partial_z \phi|^2+ m^2 |\phi|^2\right) \, .
\end{eqnarray}
Here we put $\partial_\mu \phi=0$ which is satisfied for the solution.
The $m$ dependence can be absorbed 
into the rescaling $z_{\rm new}= m z$, such that
\begin{eqnarray}
S= -  m \int d^dx \int dz_{\rm new}\;  F(|\phi|^2)\left(
|\partial_z^{\rm new} \phi|^2  + |\phi|^2\right) \, .
\end{eqnarray}
In this expression the derivative $\partial_z^{\rm new}$ 
is with respect to $z_{\rm new}$. 
Now the $m$ dependence appears only as an overall factor, so the equation of motion
is independent of $m$. Then the domain wall solution is written as 
$\phi_0=f(z_{\rm new})$ which is independent of $m$. Recovering the $m$ dependence, we obtain a solution $\phi_0 = f(m z)$. Substituting this into the action, we obtain an on-shell action 
\begin{eqnarray}
S_{\rm dw} =  -m \int d^dx 
\left[
\int dz\;  F(|\phi|^2)\left(|\partial_z \phi|^2  + |\phi|^2\right)
\biggm|_{\phi = f(z)}
\right]
\end{eqnarray}
where the last factor written with ``[ ]" is independent of $m$. Denoting it as $-A$,
we obtain a linear dependence,
\begin{eqnarray}
g(m) = A m\, .
\end{eqnarray}

So, in summary, for a $(d+1)$-dimensional system whose action is given by
\begin{eqnarray}
S= -\int d^{d+1}x \;  F(|\phi|^2)\left(|\partial_\mu \phi|^2  +|\partial_z \phi|^2
+ m^2 |\phi|^2\right)  \, ,
\label{S4V0}
\end{eqnarray}
the internal moduli appearing in the solution as
\begin{eqnarray}
\phi = e^{i m\epsilon(x^\mu)} f(mz)
\label{me}
\end{eqnarray}
has an effective action of the Nambu-Goto form,
\begin{eqnarray}
S_{\rm dw} = \int d^dx \; A m \sqrt{1 + (\partial_\mu \epsilon)^2} \, ,
\label{dwNG}
\end{eqnarray}
where the overall coefficient $A$ is given by
\begin{eqnarray}
A \equiv 
- \int dz\;  F(|\phi|^2)\left(|\partial_z \phi|^2  + |\phi|^2\right)
\biggm|_{\phi = f(z)}\, .
\end{eqnarray}

The Nambu-Goto action in 1+2 dimensions is equivalent to the Born-Infeld
action. So we complete the derivation of the Born-Infeld action as an effective
action of an internal moduli of a domain wall, for a class of 1+3 dimensional
complex scalar field theories
whose Lagrangian is of the form (\ref{S4V0}).

\subsection{The speed limit in internal space}

The Nambu-Goto action (\ref{dwNG}) shows the existence of the speed limit.
As we parameterized the internal space as (\ref{me}), this shows that
the speed limit in the internal space is $m$. 
So, the mass term in the original action has quite an important property:
it serves as the speed limit in the internal space.

Looking at the generic effective action (\ref{dwres}) which we obtained,
it exhibits interesting structure: it is a function of a Nambu-Goto Lagrangian.
Since the Nambu-Goto Lagrangian indicates the speed limit, the
generic action may have the speed limit in the internal space.
The critical speed is indeed the mass, $m$.

One may notice that the separation between the mass term and 
the potential term $V$ is arbitrary in our calculation. Indeed, one can 
split the original mass term as
$m^2 |\phi|^2 = (m_1)^2 |\phi|^2 + (m_2)^2|\phi|^2$
where $m = \sqrt{(m_1)^2 + (m_2)^2}$, and regard the $(m_2)^2$ term
as a potential term, while $(m_1)^2$ term as a mass term. 
Following the same procedure as the previous subsection, we obtain
an effective action
\begin{eqnarray}
S_{\rm dw}
= \int d^dx \; h\left(m_1 \sqrt{1 + (\partial_\mu \eta)^2}, m_2\right) 
\label{eff12}
\end{eqnarray}
where
\begin{eqnarray}
h(m_1,m_2) \equiv \int dz \, {\cal L} \biggm|_{\phi = \phi_0(z)} \, .
\end{eqnarray}
Here the internal moduli field $\eta$ is defined as
\begin{eqnarray}
\phi = e^{i m_1\eta(x^\mu)} f(mz) \, .
\label{me2}
\end{eqnarray}
Notice the difference from (\ref{me}): the definitions of the moduli fields 
are related as
\begin{eqnarray}
m_1\eta(x^\mu) = m \epsilon(x^\mu) \, .
\label{me3}
\end{eqnarray}
Now, obviously the splitting of the mass term should
not change the resultant effective action, so (\ref{eff12}) should be equal to
(\ref{dwres}). Indeed, if one notice the equality
\begin{eqnarray}
h(m_1,m_2) = g(m)\bigm|_{m = \sqrt{(m_1)^2 + (m_2)^2}} \, ,
\end{eqnarray}
it is easy to show the equivalence of (\ref{eff12}) and 
(\ref{dwres}), through the relation (\ref{me3}).
%
%

\section{Examples of effective actions on domain walls}

\subsection{Massive ${\mathbb C}P^1$ sigma model}

An example satisfying this condition $V=0$ and also the condition of
having two minima in (\ref{Vcond})
is a massive ${\mathbb C}P^1$ 
sigma model (\ref{cp1}), 
\begin{eqnarray}
F(|\phi|^2) = \frac{1}{(1 + |\phi|^2)^2} \, , \quad V = 0 \, .
\end{eqnarray}
The vacua are located at $\phi=0$ and $\phi=\infty$ as mentioned above, 
and an explicit domain wall solution is
\begin{eqnarray}
\phi_0(z) = e^{m z}\, .
\end{eqnarray}
Since the system is with $V=0$, the solution is of the form $f(mz)$ as explained
in the previous section. The on-shell action is given by
\begin{eqnarray}
S_{\rm dw} =  -m \int d^dx 
\left[
\int dz\;  \frac{|\partial_z \phi|^2  + |\phi|^2}{(1 + |\phi|^2)^2}
\biggm|_{\phi = e^{z}}
\right]
= \int d^dx \ ( -m ) \, .
\end{eqnarray}
Thus the effective action is of the Nambu-Goto type given in Eq.~(\ref{dwNG}) 
with $A= -1$.

Let us verify if the Nambu-Goto action correctly describe dynamics of domain wall. For that purpose, it is simple
to see a time-dependent solution, namely the so-called Q-kink domain wall \cite{Abraham:1992vb}. The solution is obtained through 
a standard Bogomol'nyi technique as
\begin{eqnarray}
M &=& m \int dz\ \frac{|\dot \phi - i \phi \sin \alpha|^2 + |\phi'  - \phi \cos \alpha|^2 - i  (\dot\phi\phi^*-\phi\dot\phi^*) \sin\alpha 
+ (|\phi|^2)'\cos\alpha}{(1+|\phi|^2)^2} \nonumber\\
&\ge& m T \cos\alpha + m Q \sin\alpha,
\end{eqnarray}
with the Noether charge $Q = \int dz\ \frac{ - i  (\dot\phi\phi^*-\phi\dot\phi^*) }{(1+|\phi|^2)^2}$
and the topological charge $T= \int dz\ \frac{ (|\phi|^2)'}{(1+|\phi|^2)^2}$.
The energy bound from below is the most stringent when $\tan\alpha = Q/T$.
Expressing $\sin\alpha = \omega/m$ and $\cos\alpha = \sqrt{1-(\omega/m)^2}$, we get the BPS solution and mass formula 
of the Q-kink domain wall
\begin{eqnarray}
\phi = e^{i\tilde \omega x^0 + \sqrt{1-\tilde\omega^2}\, z},\quad
M = m \sqrt{T^2 + Q^2} = \frac{m}{\sqrt{1 - \tilde \omega^2}},\quad \tilde\omega = \frac{\omega}{m}.
\label{eq:sol_qkdw}
\end{eqnarray}
We can derive the same mass formula from the effective Lagrangian of the Nambu-Goto type.
The Lagrangian is 
\begin{eqnarray}
L_{\rm NG} = - m \sqrt{1 + (\partial_\mu \epsilon)^2}\,.
\end{eqnarray}
A conjugate momentum is given by $\pi_\epsilon = \frac{\delta L_{\rm NG}}{\delta \partial_0\epsilon} 
= - m \partial^0\epsilon/\sqrt{1 + (\partial_\mu \epsilon)^2}$. Then, the Hamiltonian takes the form 
\begin{eqnarray}
H_{\rm NG} = (\p_0\epsilon) \pi_\epsilon - L_{\rm NG} 
= m \frac{1+(\partial_i\epsilon)^2}{\sqrt{1+(\partial_\mu\epsilon)^2}},\quad (i = 1,2).
\end{eqnarray}
Reading $x^0$ dependence of $\epsilon(x^\mu)$ from the Q-kink domain wall solution (\ref{eq:sol_qkdw}),
we find $\epsilon(x^\mu) = \tilde \omega x^0$. Plugging this into $H_{\rm NG}$, we find 
\begin{eqnarray}
H_{\rm NG} = \frac{m}{\sqrt{1-\tilde\omega^2}}.
\end{eqnarray}
This is precisely the same as the BPS mass formula given in Eq.~(\ref{eq:sol_qkdw}).
Thus, we confirm the Nambu-Goto action works very well as the effective action.

\subsection{An additional potential to massive ${\mathbb C}P^1$ sigma model: A model}

%
\begin{figure} 
\includegraphics[width=10cm]{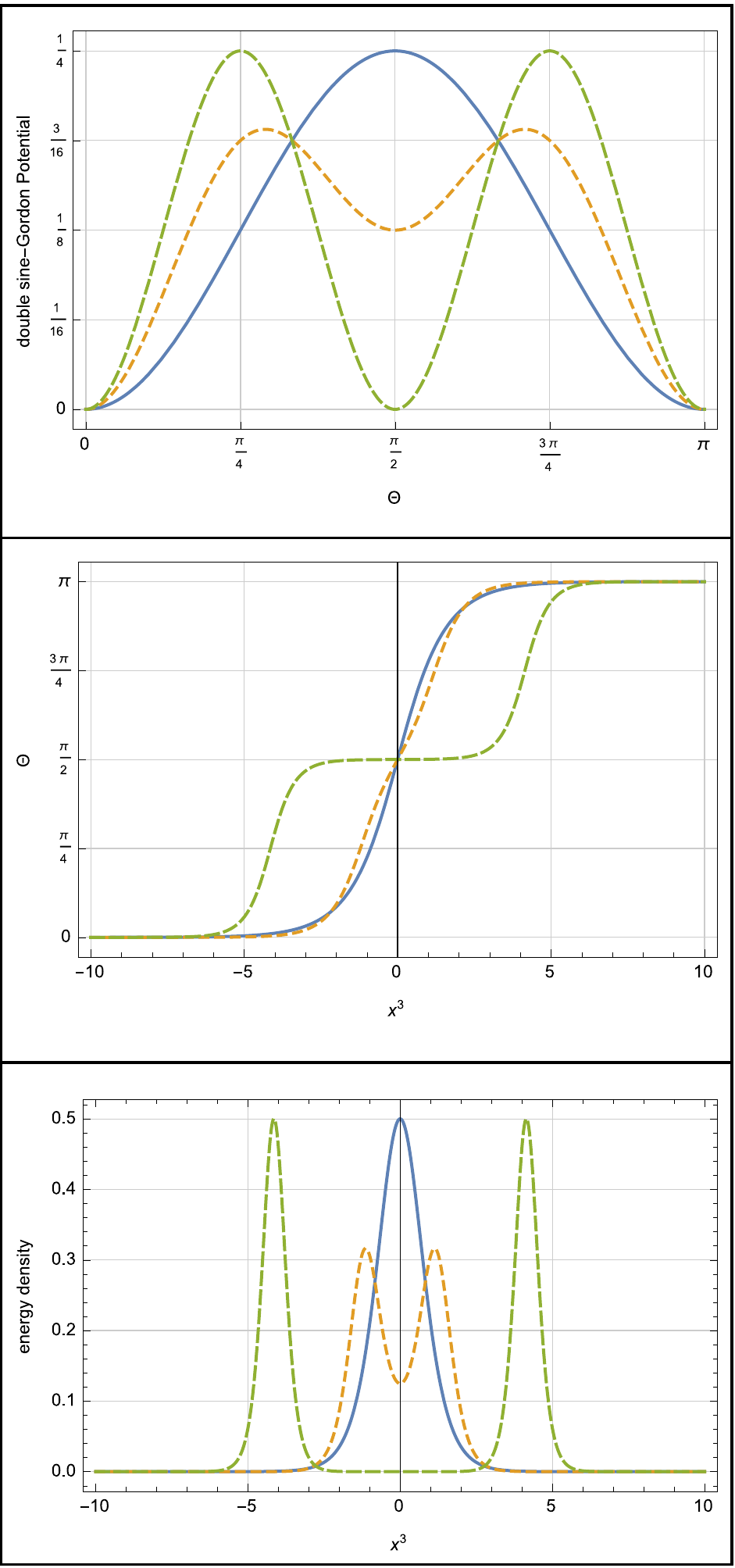}
\caption{The upper, middle, and lower panels show the double sine-Gordon potential, 
the domain wall solutions, and energy densities, respectively.
Blue (solid), yellow (dashed), and green (long-dashed)
correspond to $(m,\lambda) =(1,0),\ (1/2,1/2),\ (10^{-3},1)$.}
\label{Fig:double_sg}
\end{figure}
%

In the previous subsection, we have seen the simplest example in which the effective action of the
domain wall corresponds to the usual Nambu-Goto action.
In this section, we will see an example that the effective theory is not of the simple
Nambu-Goto type. 

We will again consider the massive $\mathbb{C}P^1$ model. 
But we introduce an additional higher order interaction term (\ref{V_A})
\begin{eqnarray}
{\cal L} &=& \frac{|\partial_M \phi|^2 - m^2 |\phi|^2}{(1+|\phi|^2)^2} - V, 
\label{eq:L_dCP1}\\
V&=&  \frac{4\lambda |\phi|^2(1-|\phi|^2)^2}{(1+|\phi|^2)^4},
\label{eq:pot_dsG}
\end{eqnarray}
where we assume $\lambda \ge 0$. The total scalar potential is positive definite, so that
$\phi = 0$ and $\phi = \infty$ remain as
the global vacua with zero vacuum energy.  
In addition, a new local minimum appears at $|\phi| = 1$ due to the additional term in the potential.
Thus, a domain wall interpolating the vacua $\phi = 0,\infty$ still exists but it is deformed 
compared to the one in the massive $\mathbb{C}P^1$ sigma model in the previous subsection. 
An advantage of the particular choice of the potential (\ref{eq:pot_dsG}) is that we will have an analytic 
solution of the domain wall with which we can analytically obtain an effective action of the deformed domain wall.

In order to illustrate the situation better, 
let us rewrite the Lagrangian in terms of a spherical coordinate 
\begin{eqnarray}
\phi = e^{i\Phi}\tan \frac{\Theta}{2}.
\label{eq:chco_cp_sG}
\end{eqnarray}
Then, the Lagrangian is written in the following form
\begin{eqnarray}
{\cal L} = \frac{1}{4}\left[(\partial_M\Theta)^2 +\sin^2\Theta (\partial_M\Phi)^2 - m^2 \sin^2\Theta -  \lambda \sin^2 2\Theta\right].
\label{eq:lag_modelA}
\end{eqnarray}
The scalar potential in this Lagrangian is identical to the so-called double sine-Gordon potential, see Fig.~\ref{Fig:double_sg}. 
The double sine-Gordon model has been studied for a long time, and a domain wall solution is known to be
\begin{eqnarray}
\Theta = \arccos\left[ \mp \frac{ \sinh \left( m \sqrt{1+\gamma} (z-Z)\right)}{\sqrt{
\cosh^2\left( m \sqrt{1+\gamma} (z-Z)\right) + \gamma}}\right],\quad \Phi = \epsilon,\quad
\gamma = \frac{4\lambda}{m^2},
\end{eqnarray}
where $Z$ and $\epsilon$ are constants.
For the upper (lower) sign, $\Theta$ goes to $0$ ($\pi$) as $z \to -\infty$ and to $\pi$ ($0$) as $z \to + \infty$.

Now, we can easily translate the above domain wall solution in terms of the original $\mathbb{C}P^1$ field $\phi$.
It is of a bit complicated form
\begin{eqnarray}
\phi_0 = e^{i\epsilon}
\left[\frac{\sqrt{\gamma + \cosh ^2\left(m\sqrt{1+\gamma}(z-Z)\right)} 
\pm  \sinh \left(m\sqrt{1+\gamma} (z-Z)\right)}{\sqrt{\gamma + \cosh ^2\left(m\sqrt{1 + \gamma} (z-Z)\right)}
\mp  \sinh \left(m\sqrt{1 + \gamma} (z-Z)\right)}
\right]^{\frac{1}{2}}.
\label{eq:sol_dsG_phi}
\end{eqnarray}
For the upper (lower) sign, $\phi$ goes to $0$ ($\infty$) as $z \to -\infty$ and to $\infty$ ($0$) as $z \to + \infty$.
Let us obtain the mass of the domain wall. It can simply be done by making use of the standard Bogomol'nyi technique as
\begin{eqnarray}
M 
&=& \int dz\ \frac{1}{(1+|\phi|^2)^2} \Bigg\{\ \left|\phi' \mp e^{i\epsilon} 
m\sqrt{ |\phi|^2 + 
\frac{\gamma|\phi|^2(1-|\phi|^2)^2}{(1+|\phi|^2)^2}}\right|^2 \nonumber\\
&\pm& \left(e^{-i\epsilon} \phi' + e^{i\epsilon} \bar\phi'\right)  
m\sqrt{ |\phi|^2 + 
\frac{\gamma|\phi|^2(1-|\phi|^2)^2}{(1+|\phi|^2)^2}}\ \Bigg\}\nonumber\\
&\ge& \pm m \int dz\ \frac{\left(e^{-i\epsilon} \phi' + e^{i\epsilon} \bar\phi'\right)  }{(1+|\phi|^2)^2} 
\sqrt{|\phi|^2 + 
\frac{\gamma |\phi|^2(1-|\phi|^2)^2}{(1+|\phi|^2)^2}},
\end{eqnarray}
where $\epsilon$ is an arbitrary real constant.
The Bogomol'nyi bound is saturated when the following first order differential equation is satisfied
\begin{eqnarray}
\phi' = \pm m e^{i\epsilon} 
\sqrt{ |\phi|^2 + 
\frac{\gamma |\phi|^2(1-|\phi|^2)^2}{(1+|\phi|^2)^2}}.
\end{eqnarray}
This is indeed solved by $\phi$ given in Eq.~(\ref{eq:sol_dsG_phi}).
Thus, the mass is given by
\begin{eqnarray}
M\left(m,\gamma \right) &=&m  \int_0^\infty df\ \frac{2}{(1+f^2)^2} 
\sqrt{f^2 + 
\frac{\gamma f^2(1-f^2)^2}{(1+f^2)^2}} \nonumber\\
&=& \frac{1}{2} \left[\sqrt{m^2 + 4\lambda} + \frac{m^2}{\sqrt{4\lambda}} \tanh^{-1} \sqrt{\frac{4\lambda}{m^2}}\right].
\end{eqnarray}
This is, of course, identical to the mass formula of the double sine-Gordon domain wall known in the literature.

We now extend the above solution to have a $Q$-charge, which is a new solution existing only in 
the deformed massive $\mathbb{C}P^1$ model.
In order to get the solution, the Bogomol'nyi technique is again useful,
\begin{eqnarray}
M 
&=& \int \frac{dz}{(1+|\phi|^2)^2} \Bigg\{\ \left|\dot\phi - i\omega \phi\right|^2 + \left|\phi' \mp e^{i\epsilon(t)} 
\tilde m \sqrt{ |\phi|^2 + 
\frac{\tilde\gamma|\phi|^2(1-|\phi|^2)^2}{(1+|\phi|^2)^2}}\right|^2 \nonumber\\
&\pm& \tilde m \left(e^{-i\epsilon(t)} \phi' + e^{i\epsilon(t)} \bar\phi'\right)  
\sqrt{ |\phi|^2 + 
\frac{\tilde\gamma|\phi|^2(1-|\phi|^2)^2}{(1+|\phi|^2)^2}}
- i \omega (\dot\phi \bar\phi - \phi \dot{\bar\phi}) \ \Bigg\}\nonumber\\
&\ge& \int dz\Bigg\{ \pm \tilde m \frac{\left(e^{-i\epsilon(t)} \phi' + e^{i\epsilon(t)} \bar\phi'\right)  }{(1+|\phi|^2)^2} 
\sqrt{|\phi|^2 + 
\frac{\tilde\gamma |\phi|^2(1-|\phi|^2)^2}{(1+|\phi|^2)^2}}
- \frac{i\omega(\dot\phi\bar\phi-\phi\dot{\bar\phi})}{(1+|\phi|^2)^2}\Bigg\},\nonumber\\
\end{eqnarray}
where we have introduced
\begin{eqnarray}
\tilde m^2 = m^2 - \omega^2,\quad \tilde \gamma = \frac{4\lambda}{m^2 - \omega^2}.
\end{eqnarray}
The energy bound is saturated when the first order differential equations are satisfied
\begin{eqnarray}
\dot\phi = i \omega \phi,\quad
\phi' \pm e^{i\epsilon(t)} 
\tilde m \sqrt{ |\phi|^2 + 
\frac{\tilde\gamma|\phi|^2(1-|\phi|^2)^2}{(1+|\phi|^2)^2}}=0.
\end{eqnarray}
This is solved by
\begin{eqnarray}
\phi_Q = e^{i\omega t + i \epsilon_0} 
\left[\frac{\sqrt{\tilde\gamma + \cosh ^2\left(\tilde m\sqrt{1+\tilde\gamma}(z-Z)\right)} 
\pm  \sinh \left(\tilde m\sqrt{1+\tilde\gamma} (z-Z)\right)}{\sqrt{\tilde\gamma + \cosh ^2\left(\tilde m\sqrt{1 + \tilde\gamma} (z-Z)\right)}
\mp  \sinh \left(\tilde m\sqrt{1 + \tilde\gamma} (z-Z)\right)}
\right]^{\frac{1}{2}}.
\label{eq:q_sol}
\end{eqnarray}
The mass of the Q-double sine-Gordon domain wall is given by
\begin{eqnarray}
M_Q(m,\lambda,\omega) = \frac{1}{2}\left[
\sqrt{m^2-\omega^2 + 4\lambda} 
+ \frac{m^2+\omega^2}{\sqrt{4\lambda}} \tanh^{-1}\sqrt{\frac{4\lambda}{m^2-\omega^2+4\lambda}}\right].
\label{eq:M_Q}
\end{eqnarray}
Note that this corresponds to the mass of the usual $\mathbb{C}P^1$ Q domain wall in the limit $\lambda \to 0$ as
\begin{eqnarray}
M_Q(m,0,\omega) = \sqrt{m^2-\omega^2} + \frac{\omega^2}{\sqrt{m^2-\omega^2}} = \frac{m^2}{\sqrt{m^2-\omega^2}}.
\end{eqnarray}
The Q-charge density is given by
\begin{eqnarray}
Q &=& \int dx^3\ \frac{i(\phi \dot\phi^* - \dot\phi \phi^*)}{(1+|\phi|^2)^2} \nonumber\\
&=& \left[\frac{\omega}{4 \sqrt{\lambda }}  \tanh ^{-1}\left(
\frac{2 \sqrt{\lambda }}{\sqrt{4 \lambda +m^2-\omega ^2}} \tanh \left(x \sqrt{4 \lambda +m^2-\omega ^2}\right)
\right)
\right]^{x^3=+\infty}_{x^3=-\infty} \nonumber\\
&=& \frac{\omega}{2 \sqrt{\lambda }}  \tanh ^{-1}\frac{2 \sqrt{\lambda }}{\sqrt{4 \lambda +m^2-\omega ^2}}.
\label{eq:Q_modelA}
\end{eqnarray}

Let us next compare the values of Lagrangian for the static and Q domain wall solutions. 
Substituting $\phi_0$ in Eq.~(\ref{eq:sol_dsG_phi}) and $\phi_Q$ in Eq.~(\ref{eq:q_sol}) into
the Lagrangian (\ref{eq:L_dCP1}) and integrating it over $z$, we get
\begin{eqnarray}
L [\phi_0] &=& -\frac{1}{2}\left[
\sqrt{m^2 + 4\lambda} 
+ \frac{m^2}{\sqrt{4\lambda}} \tanh^{-1}\sqrt{\frac{4\lambda}{m^2+4\lambda}}\right],
\label{eq:eff_lag_def_cp1}\\
L [\phi_Q] &=& -\frac{1}{2}\left[
\sqrt{m^2-\omega^2 + 4\lambda} 
+ \frac{m^2-\omega^2}{\sqrt{4\lambda}} \tanh^{-1}\sqrt{\frac{4\lambda}{m^2-\omega^2+4\lambda}}\right].
\end{eqnarray} 
The latter can be obtained by just replacing $m$ by $\sqrt{m^2 - \omega^2}$ in the former. Having
$\Lambda^\alpha{}_\alpha= \frac{1}{\sqrt{1-\frac{\omega^2}{m^2}}}$ as a Lorentz boost toward the $(d+2)$-th direction, 
this replacement
can be understood as exchanging $m \to m/\Lambda^\alpha{}_\alpha$. Thus, we have verified that the deformed $\mathbb{C}P^1$
model is in the category to which our prescription can apply.

Now, the effective action of the domain wall can be obtained by just replacing
\begin{eqnarray}
m \to L_{\rm NG} = m \sqrt{1 + (\p_\mu \epsilon)^2},
\end{eqnarray}
in the Lagrangian $L[\phi_0]$ given in Eq.~(\ref{eq:eff_lag_def_cp1}),
\begin{eqnarray}
L_{\rm eff} =
-\frac{1}{2}\Bigg[
\sqrt{L_{\rm NG}^2+ 4\lambda} 
+ \frac{L_{\rm NG}^2}{\sqrt{4\lambda}} \tanh^{-1}\sqrt{\frac{4\lambda}{L_{\rm NG}^2+4\lambda}}\,\Bigg].
\label{eq:eff_Lag_modelA}
\end{eqnarray}
This is very different from the standard Nambu-Goto Lagrangian ($\lambda \to 0$ limit is the Nambu-Goto Lagrangian).
Conjugate momentum is given by
\begin{eqnarray}
\Pi_\epsilon = \frac{m^2}{\sqrt{4\lambda}} \tanh^{-1}
\sqrt{\frac{4\lambda}{m^2\left(1 + (\p_\mu \epsilon)^2\right)+4\lambda}}\ \p_0\epsilon.
\end{eqnarray}
Then, the Hamiltonian is
\begin{eqnarray}
H_{\rm eff} = \frac{1}{2}\Bigg[
\sqrt{m^2\left(1 + (\p_\mu \epsilon)^2\right) + 4\lambda} 
+ \frac{m^2\left(1 + (\p_0 \epsilon)^2 + (\p_i \epsilon)^2\right)}{\sqrt{4\lambda}} \tanh^{-1}\sqrt{\frac{4\lambda}{m^2\left(1 + (\p_\mu \epsilon)^2\right)+4\lambda}}\,\Bigg].
\end{eqnarray}
Finally, we consider the solution in the effective theory
\begin{eqnarray}
\epsilon(x^\mu) = \frac{\omega}{m}t + \epsilon_0.
\label{eq:eff_sol}
\end{eqnarray}
The corresponding energy density is 
\begin{eqnarray}
H_{\rm eff}
= \frac{1}{2}\left[
\sqrt{m^2-\omega^2 + 4\lambda} 
+ \frac{m^2+\omega^2}{\sqrt{4\lambda}} \tanh^{-1}\sqrt{\frac{4\lambda}{m^2-\omega^2+4\lambda}}\right].
\end{eqnarray}
This is nothing but $M_Q(m,\lambda,\omega)$ given in Eq.~(\ref{eq:M_Q}).
Furthermore, the Noether charge of the shift symmetry $\epsilon \to \epsilon + \delta$ for the solution (\ref{eq:eff_sol})  is given by
\begin{eqnarray}
Q_{\rm eff}\bigg|_{\epsilon_Q} = \frac{\partial L}{\partial (m \dot\epsilon)}\bigg|_{\epsilon_Q} = \frac{1}{m} \Pi_\epsilon\bigg|_{\epsilon_Q} 
= \frac{\omega}{2\sqrt{\lambda}} \tanh^{-1}
\sqrt{\frac{4\lambda}{m^2- \omega^2+4\lambda}}.
\end{eqnarray}
This Noether charge in the effective theory is exactly the same as that in the original theory, see Eq.~(\ref{eq:Q_modelA}).
Thus the effective Lagrangian (\ref{eq:eff_Lag_modelA}) correctly reproduce not only the mass but also the conserved charge.

\subsection{An additional potential to massive ${\mathbb C}P^1$ sigma model: B model}

%
\begin{figure} 
\includegraphics[width=10cm]{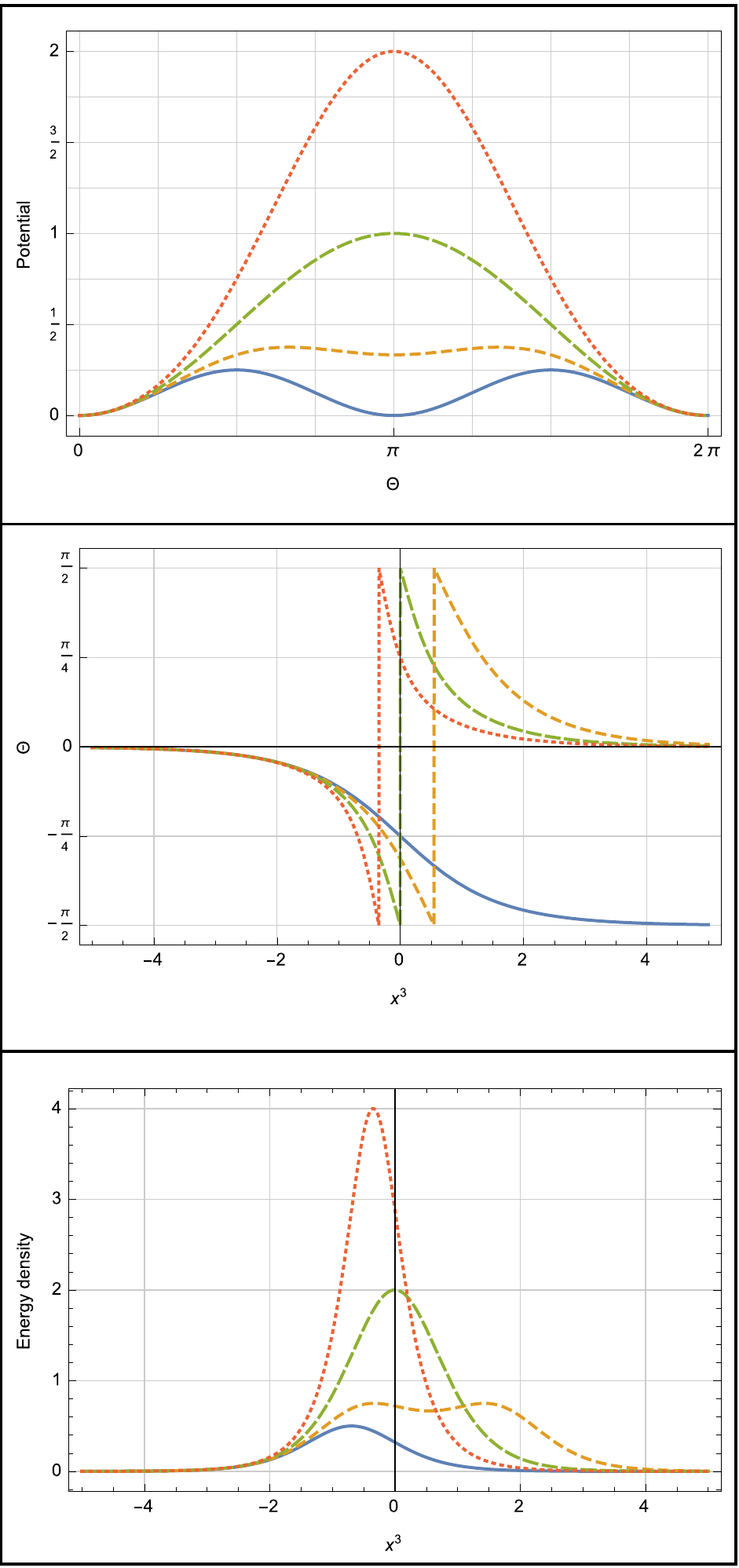}
\caption{The upper, middle, and lower panels show the double sine-Gordon potential B,  
the domain wall solutions, and energy densities, respectively.
Blue (solid), yellow (dashed), green (long-dashed), and orange (short-dashed)
correspond to $(m,\eta) =(1,0),\ (1,1/3),\ (1,1)$, and $(1,2)$.}
\label{Fig:double_sg_B}
\end{figure}
%

Let us next consider a different scalar potential from Eq.~(\ref{eq:pot_dsG}) for the $\mathbb{C}P^1$ model
\begin{eqnarray}
V = \frac{\eta |\phi|^4}{(1+|\phi|^2)^2},\quad (\eta > 0).
\end{eqnarray}
This model is not explained in the main part of the paper, but we consider this model to give
further support to the main result of the paper.
This potential lifts the vacuum at the south pole $|\phi| = \infty$ while the point $\phi = 0$ being left as the unique vacuum.
As shown in Fig.~\ref{Fig:double_sg_B}, the vacuum $|\phi| = \infty$ remains as a local minimum for $m^2 > \eta$ while
it becomes a global maximum for $m^2 \le \eta$.
In terms of the spherical coordinate (\ref{eq:chco_cp_sG}), we find the Lagrangian in the following expression
\begin{eqnarray}
{\cal L} = \frac{1}{4}\left[(\partial_M\Theta)^2 +\sin^2\Theta (\partial_M\Phi)^2 - m^2 \sin^2\Theta -  4\eta \sin^2 \frac{\Theta}{2}\right].
\end{eqnarray}
This is quite similar to the model (\ref{eq:lag_modelA}). A difference is period of the scalar potential in the $\Theta$ direction.
A static domain wall solution of the equations of motion is given by
\begin{eqnarray}
\phi_0 = e^{i\epsilon} \frac{2m^2 e^{m (z-Z)}}{m^2 - \eta e^{2m (z-Z)}}.
\label{eq:sol_dw_modelB}
\end{eqnarray}
One should change coordinate from $\phi$ to $\phi' = 1/\phi$ in the vicinity of the point $\phi_0 \to \infty$.
This solution interpolates the north pole ($\phi = 0$) at $z \to -\infty$, and passing through the south pole ($|\phi|=\infty$),
it reaches back to the north pole. Since the field trajectory goes across the potential hill twice, the configuration is again 
a bound state of two domain walls at a finite distance. This solution might be unstable against generating tachyonic fluctuations because
the $S^1$ trajectory surrounding  the $S^2$ target space is contractible.
In the following, we will not worry about the tachyonic modes at all. Instead, we will concentrate on the zero modes and massive modes
around the background solution given in Eq.~(\ref{eq:sol_dw_modelB}), since our main interest of this paper is put in the effective theory of the
zero modes.

Let us start with giving a Q-extension of the static domain wall solution in Eq.~(\ref{eq:sol_dw_modelB}),
\begin{eqnarray}
\phi_Q = e^{i\omega t + \epsilon_0} \frac{2(m^2-\omega^2) e^{\sqrt{m^2-\omega^2} (z-Z)}}{m^2-\omega^2 - \eta e^{2\sqrt{m^2-\omega^2} (z-Z)}}.
\end{eqnarray}
We assume $\omega^2 < m^2$ because $\omega^2 = m^2$ corresponds to the vacuum solution, and
$\omega^2 > m^2$ does not solve the equation of motion.
The tension of this solution is given in a bit complicated form as 
\begin{eqnarray}
M_Q^{m^2 - \eta - \omega^2 >0} &=& \frac{2  \left(m^2-\eta \right) \sqrt{m^2-\omega^2}}{m^2-\eta -\omega^2} \nonumber\\
&+& \frac{\eta  \left(m^2 - \eta -2 \omega ^2\right) }{2 \left(m^2-\eta -\omega ^2\right)^{3/2}}
\log \left(\frac{ 2m^2 - \eta -2 \omega ^2 + 2  \sqrt{(m^2-\omega ^2)(m^2 -\eta -\omega ^2)}}{2 m^2 - \eta -2 \omega ^2 - 2 \sqrt{(m^2-\omega ^2)(m^2-\eta-\omega ^2)}}\right),\\
M_Q^{m^2 - \eta - \omega^2 = 0} &=& \frac{4 m^2}{\sqrt{m^2-\omega ^2}},\\
M_Q^{m^2 - \eta - \omega^2 < 0} &=& 
\frac{4 \left(\eta -m^2\right) \sqrt{\left(m^2-\omega ^2\right) \left(\eta -m^2+\omega ^2\right)}+\pi  \eta  \left(\eta -m^2+2 \omega ^2\right)}{2 \left(\eta -m^2+\omega ^2\right)^{3/2}} \nonumber\\
&+& \frac{\eta  \left(\eta -m^2+2 \omega ^2\right) }{\left(\eta -m^2+\omega ^2\right)^{3/2}}
\cot ^{-1}\left(\frac{2 \sqrt{\left(m^2-\omega ^2\right) \left(\eta -m^2+\omega ^2\right)}}{\eta -2 m^2+2 \omega ^2}\right).
\label{eq:M_Q_modelB}
\end{eqnarray}
As is done in the previous subsection, our first non-trivial check is comparing the value of integration of the Lagrangian 
over the transverse coordinate $z$ for
the static solution $\phi_0$ and $\phi_Q$:
\begin{eqnarray}
L[\phi_0]^{m^2 - \eta > 0} &=& -2 m +  \frac{\eta  }{2 \sqrt{m^2-\eta }}
\log \left(\frac{m \left(m-\sqrt{m^2-\eta }\right)+m^2-\eta }{m \left(m + \sqrt{m^2-\eta }\right)+ m^2-\eta }\right),\\
L[\phi_0]^{m^2 - \eta = 0} &=& -4m,\\
L[\phi_0]^{m^2 - \eta < 0} &=& -2 m -\frac{\eta  }{ \sqrt{\eta -m^2}}
\left(\frac{\pi}{2} - \tan ^{-1}\left(\frac{2 m^2-\eta }{2 m \sqrt{\eta -m^2}}\right)\right) .
\end{eqnarray}
As is expected, we find that $L[\phi_Q]$ is obtained by just replacing $m$ by $\sqrt{m^2-\omega^2}$ in $L[\phi_0]$.
Now, we can construct a low energy effective theory by replacing $m$ with $L_{\rm NG} = m\sqrt{1+(\partial_\mu \epsilon)^2}$.
Then, the Hamiltonian in the effective theory is obtained by a standard way. It for $\epsilon$ dependent only on $x^0$ is 
expressed as
\begin{eqnarray}
H_{\rm eff}^{m^2 - \eta - \dot\epsilon^2 >0} &=& \frac{2  \left(m^2-\eta \right) m\sqrt{1-\dot\epsilon^2}}{m^2(1-\dot\epsilon^2)-\eta} \nonumber\\
&+& \frac{\eta  \left(m^2(1-2\dot\epsilon^2)-\eta\right) }{2 \left(m^2(1-\dot\epsilon^2)-\eta\right)^{3/2}}
\log \left(\frac{ 2m^2(1-\dot\epsilon^2)-\eta + 2  m \sqrt{(1-\dot\epsilon^2)(m^2(1-\dot\epsilon^2)-\eta)}}{2m^2(1-\dot\epsilon^2)-\eta - 2  m \sqrt{(1-\dot\epsilon^2)(m^2(1-\dot\epsilon^2)-\eta)}}\right),\\
H_{\rm eff}^{m^2 - \eta - \dot\epsilon^2 = 0} &=& \frac{4 m}{\sqrt{1-\dot\epsilon^2}},\\
H_{\rm eff}^{m^2 - \eta - \dot\epsilon^2 < 0} &=& 
\frac{4 \left(\eta -m^2\right) m\sqrt{\left(1-\dot\epsilon^2\right) \left(\eta -m^2(1-\dot\epsilon^2)\right)}+\pi  \eta  \left(\eta -m^2(1-2\dot\epsilon^2)\right)}{2 \left(\eta -m^2(1-\dot\epsilon^2)\right)^{3/2}} \nonumber\\
&+& \frac{\eta  \left(\eta -m^2(1-2\dot\epsilon^2)\right) }{\left(\eta -m^2(1-\dot\epsilon^2)\right)^{3/2}}
\cot ^{-1}\left(\frac{2 m \sqrt{\left(1-\dot\epsilon^2\right) \left(\eta -m^2(1-\dot\epsilon^2)\right)}}{\eta -2m^2 (1-\dot\epsilon^2)}\right).
\end{eqnarray}
By putting a solution $\epsilon = \omega t/m$ in the effective theory, we exactly reproduce the mass formula
(\ref{eq:M_Q_modelB}) in the original non-linear sigma model.

\begin{figure} 
\includegraphics[width=10cm]{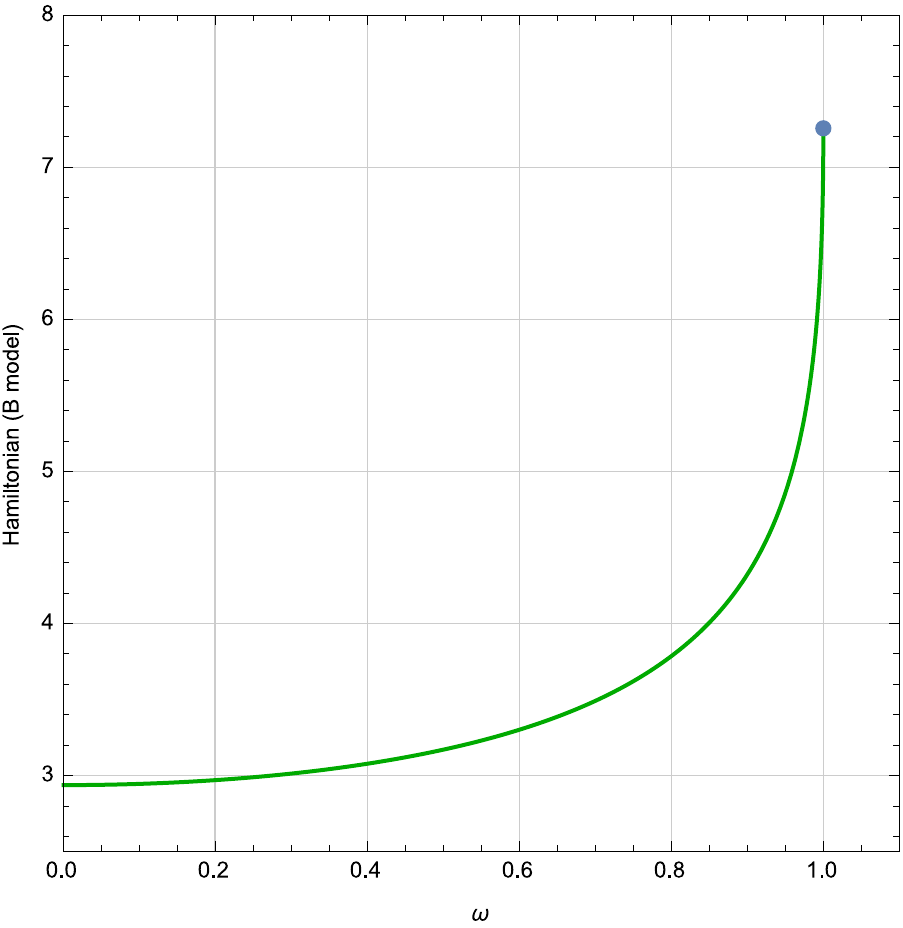}
\caption{The on-shell Hamiltonian for $m=1$ and $\eta=1/3$. 
$H = 2-\frac{\log \left(49-20 \sqrt{6}\right)}{2 \sqrt{6}}$ at $\omega = 0$ and
$H= 4\pi/\sqrt{3}$ at $\omega = 1$.}
\label{Fig:H_B}
\end{figure}

A peculiar feature of the B model is that the mass does not diverge at $\omega/m = 1$.
Namely, there is no speed limit in the internal moduli. An intuitive explanation is the following.
As $\omega \to m$, the effective mass becomes small. In the B model, the potential of the small period $(m^2 - \omega^2)\sin^2\Theta$
vanishes. This means that the two constituent domain walls are further confined into one large domain wall, see Fig.~\ref{Fig:double_sg_B}.
Therefore, no flattering and destroying the domain wall occurs in this model, see Fig.~\ref{Fig:H_B}.  
%


\section{Effective action for a symmetry breaking at vacuum}

We consider a generic Lagrangian of a complex scalar field $\phi(x^\mu)$ 
$(\mu=0,1,2,3)$. 
The Lagrangian is, for simplicity;
\begin{eqnarray}
S= \int d^4x \; {\cal L} \, , \quad
{\cal L} = - |\partial_\mu \phi|^2 
- m^2 |\phi|^2 - V(|\phi|^2) \, .
\label{S4d}
\end{eqnarray}
We consider a spontaneous symmetry breaking,
\begin{eqnarray}
V = c  |\phi|^2 + \lambda |\phi|^4 \, ,
\end{eqnarray}
where $-\mu^2 = m^2 + c < 0$ to make sure that there occurs
the symmetry breaking, as the original Lagrangian looks
\begin{eqnarray}
{\cal L} = - |\partial_\mu \phi|^2 
-\Bigl(-\mu^2 |\phi|^2 + \lambda |\phi|^4\Bigr) \, .
\end{eqnarray}
We treat $m$ and $c$ independently in our analysis, although 
a physically important quantity is the combination $\mu$.

A classical solution is
\begin{eqnarray}
\phi_0 = e^{i m \epsilon} \xi, \quad \epsilon \in {\mathbb R} \, .
\end{eqnarray}
Here $\xi$ is the Higgs vet and is constant, given explicitly as
\begin{eqnarray}
\xi = \sqrt{\frac{-c-m^2}{2\lambda}}\, .
\end{eqnarray}
The vacuum depends on the parameter $\epsilon$,
which can be later become a massless Nambu-Goldstone mode.

It is straightforward to obtain the on-shell action, but here we 
follow the procedures using the extra dimensions since it is instructive.
First, we upgrade the system to a $(d+2)$-dimensional system,
\begin{eqnarray}
\Phi(x^\mu,z,\alpha) = e^{im\alpha}
\sum_{n=-\infty}^{\infty} e^{in\alpha/R} \phi^{(n)}(x^\mu) \, .
\label{SSdef2}
\end{eqnarray}
The action is
\begin{eqnarray}
S &=& \frac{1}{2\pi R} \int_0^{2\pi R} \!\!\! d\alpha \; 
\int d^dx \, dz \, {\cal L}_{d+2} \, ,
\\
{\cal L}_{d+2} &=& - |\partial_M \phi|^2 - V(|\phi|^2) \, .
\label{S5}
\end{eqnarray}
The Sherk-Schwarz boundary condition is
\begin{eqnarray}
\Phi(x^\mu, \alpha + 2\pi R) = e^{2 \pi i m R }
\Phi(x^\mu, \alpha)  \, .
\end{eqnarray}

Let us consider a boosted solution in $(d+2)$ dimensions. The original solution is
\begin{eqnarray}
\Phi = \Phi_0(\alpha) \equiv e^{i m \alpha} \xi
\label{up4sol}
\end{eqnarray}
and it is boosted by a Lorentz transformation
\begin{eqnarray}
\tilde{\alpha} = 
\Lambda^\alpha_{\;\; \alpha} \alpha + \Lambda^\alpha_{\;\; \mu} x^\mu \, ,
\quad
\tilde{x}^\mu =
\Lambda^\mu_{\;\; \alpha} \alpha + \Lambda^\mu_{\;\; \nu} x^\nu \, .
\end{eqnarray}
the result is 
\begin{eqnarray}
\Phi & = & \Phi_0(\tilde{\alpha})
\nonumber \\
 &=& e^{i m \tilde{\alpha}} \xi
\nonumber \\
& = & e^{i m \Lambda^\alpha_{\;\; \alpha} \alpha}
e^{i m \Lambda^\alpha_{\;\; \mu} x^\mu} \xi \, .
\label{newsol2}
\end{eqnarray}

With this new solution, we can
calculate the effective action as before. 
With the general coordinate transformation
\begin{eqnarray}
\tilde{\alpha} = 
\Lambda^\alpha_{\;\; \alpha} \alpha + \Lambda^\alpha_{\;\; \mu} x^\mu \, ,
\quad 
\tilde{x}^\mu = 
 x^\mu \, ,
\end{eqnarray}
the substitution of the boosted solution to the action gives 
\begin{eqnarray}
S &=& 
\frac{1}{2\pi R} 
\int_{\Lambda^\alpha_{\;\; \mu} x^\mu}^{2\pi R\Lambda^\alpha_{\;\; \alpha} +\Lambda^\alpha_{\;\; \mu} x^\mu} \frac{d\tilde{\alpha}}{\Lambda^\alpha_{\;\; \alpha}} \; 
\int d^4x  \, 
\left[{\cal L}_{d+2} \biggm|_{\Phi = \Phi_0(\alpha)} \right]_{\alpha \;\; {\rm replaced
\;\; by \;\;} \tilde{\alpha}}
\nonumber
\\
& = &
\int d^4x \, {\cal L}_{d+2} \biggm|_{\Phi = \Phi_0(\alpha)} \, 
\nonumber
\\
& = &
\int d^4x \,  
\left[-m^2 \xi^2 - V(\xi^2)\right]
\nonumber
\\
& = &
\int d^4x \,  
\frac{(c+m^2)^2}{4\lambda}
\, .
\label{Sinv2}
\end{eqnarray}
Replacing $m$ by $m/\Lambda^\alpha_{\;\; \alpha}
=m\sqrt{1 + (\partial_\mu \epsilon)^2}$, 
we obtain the effective action for the internal moduli field  $\epsilon(x^\mu)$ 
as
\begin{eqnarray}
S &=& \int d^4x \, \, 
\frac{(c+m^2 + m^2 (\partial_\mu \epsilon)^2)^2}{4\lambda}
\nonumber \\
&=& \int d^4x \left[
\frac{\mu^4}{4\lambda}
 - \frac{ m^2 \mu^2}{2\lambda}
  (\partial_\mu \epsilon)^2
+ \frac{m^4}{4\lambda} (  (\partial_\mu \epsilon)^2)^2
\right]
 \, .
\label{dwres2}
\end{eqnarray}
So, if we rescale the moduli field $\epsilon(x^\mu)$ as
\begin{eqnarray}
N(x^\mu) \equiv \frac{m\mu}{\sqrt{\lambda}}\epsilon (x^\mu) \, , 
\end{eqnarray}
the effective action is
\begin{eqnarray}
S = \mbox{const.} +
\int \! d^4x \left[
-\frac12 (\partial_\mu N)^2 + \frac{\lambda}{4\mu^4}
\left((\partial_\mu N)^2\right)^2
\right] \, .
\end{eqnarray}
The normalization of the Nabmu-Goldstone field is encoded in
the original field as
\begin{eqnarray}
\phi = \frac{\sqrt{\mu^2-\frac{\lambda}{\mu^2}(\partial_\mu N)^2}}{\sqrt{2\lambda}}
\exp[i (\sqrt{\lambda}/\mu) N(x^\mu)]  \, .
\end{eqnarray}
This relation seems ill-defined when the inside of the square root becomes
negative. However, since the original scalar field is complex, there is no problem.
The effective action itself does not provide any reality constraint, thus there is no
speed limit.

\section{Effective action for 't Hooft Polyakov  monopole}

Let us turn to the case of the popular $SU(2)$ 't Hooft-Polyakov monopole, following
\cite{Tong:2005un}.
We find that the effective action for the internal moduli parameter $\epsilon(t)$
for the $SU(2)$ 't\,Hooft Polyakov monopole is given by a Nambu-Goto action,
that is, an action for a relativistic particle whose position is given by
$\epsilon(t)$.
Therefore we have a speed limit in the internal space.

We start with the $SU(2)$ Yang-Mills-Higgs theory
\begin{eqnarray}
S= -\frac{1}{g^2} \int \! d^4x
\left[
\frac14 (F_{\mu\nu}^a)^2 + \frac12 (D_\mu \phi^a)^2\right]
\end{eqnarray}
where we use the normalization of the $SU(2)$ generators as
${\rm tr} [T^a T^b] = \delta_{ab}/2$ ($a,b=1,2,3$), 
and the component expansion is $A_\mu = A_\mu^a T^a$
and $\phi = \phi^a T^a$.
The field strength and the covariant derivative are defined as
\begin{eqnarray}
&&
F_{\mu\nu} \equiv \partial_\mu A_\nu - \partial_\nu A_\mu
- i  [A_\mu, A_\nu] \, , 
\\
&&D_\mu \phi \equiv \partial_\mu \phi - i  [A_\mu, \phi].
\end{eqnarray}

The BPS equation for a monopole is given by
\begin{eqnarray}
B_i = D_i \phi\, , \quad E_i=0 \, ,
\end{eqnarray}
with $B_i \equiv (1/2) \epsilon_{ink} F_{jk}$ and 
$E_i \equiv F_{i0}$.
The  monopole solution is given as
\begin{eqnarray}
A_i^a = \epsilon_{aij} \hat{\bf r}_j(1-K(r))/r \, , \quad
A_0^a = \hat{\bf r}_a J(r)/r \, , \quad
\phi^a = \hat{\bf r}_a H(r)/r \, ,
\end{eqnarray}
with
\begin{eqnarray}
K(r) \equiv Cr/\sinh(Cr) \, , \quad
J(r) = 0 \, , \quad 
H(r) = Cr \coth(Cr) -1 \, .
\end{eqnarray}
Here $C$ is a constant parameter, and provides the Higgs VEV at
spatial infinity, $\phi^a = C \hat{\bf r}_a$.

The 't\,Hooft-Polyakov monopole has ${\mathbb R}^3\times S^1$ moduli space,
and the latter $S^1$ is the internal moduli parameter. It is generated by
an unbroken global part of the local symmetry generated by
\begin{eqnarray}
U = \exp\left[
-2 i \, \epsilon \,  \phi^a T^a / C\right]
\label{zeroU}
\end{eqnarray}
so that the periodicity for the constant 
moduli parameter $\epsilon$ is $ 0 \leq \epsilon < 2\pi$.
It is easy to see that the transformation leaves the scalar field solution intact, while
it changes the gauge field solution by
\begin{eqnarray}
\delta A_i =  \epsilon \frac{2}{C}\, D_i \phi \, .
\end{eqnarray}
This is the internal zero mode which we are interested in.

Now, we upgrade this constant internal moduli parameter $\epsilon$ to a one-dimensional field $\epsilon(t)$. Since the monopole is a point-like object,
its worldline is one-dimensional, so the moduli can depend only on time $t$.

It is important to note that,
once we consider a time-dependent $\epsilon(t)$, 
it amounts to an electric field.
In fact, the transformation
(\ref{zeroU}) generates also an electric field, 
\begin{eqnarray}
\delta A_0 = (\partial_0 \epsilon) \frac{2}{C}\phi  \, .
\label{A0move}
\end{eqnarray}
So the internal motion provides the electric field, and turns the monopole into
a dyon. 
The famous Julia-Zee dyon solution is given by
\begin{eqnarray}
&&K(r) \equiv C'r/\sinh(C'r) \, , \quad
\\
&&
J(r) = \tanh \gamma \; H(r) \, , \quad 
\label{J}
\\
&&H(r) = \cosh\gamma\left[ C'r \coth(C'r) -1\right] \, .
\end{eqnarray}
Here $C'$ is related to the previous $C$ as 
\begin{eqnarray}
C' \cosh \gamma = C 
\end{eqnarray}
such that the asymptotic value of the higgs field $\phi$ is the same as
that of the original monopole solution. 
The BPS equation for the dyon is
\begin{eqnarray}
(\cosh\gamma) B_i = D_i \phi \, , \quad
(\coth\gamma) E_i = D_i \phi \, .
\end{eqnarray}

Comparing
(\ref{J}) in this Julia-Zee dyon solution
with (\ref{A0move}), we find a relation
\begin{eqnarray}
(\partial_0 \epsilon)  \frac{2}{C}  = \tanh \gamma  \, .
\label{idente}
\end{eqnarray}
We calculate the effective action of the zero mode $\epsilon(t)$. 
It is sufficient to calculate the on-shell action of the dyon. Substituting the dyon solution
to the original action, we obtain
\begin{eqnarray}
S = - \frac{4\pi C'}{g^2} \int \! dt = 
- \frac{4\pi C}{g^2} \int \! dt \,  \frac{1}{\cosh\gamma} \, .
\end{eqnarray}
Using the relation (\ref{idente}) between $\gamma$ and the moduli field $\epsilon(t)$, 
we obtain the effective action for the internal moduli as
\begin{eqnarray}
S =  
- \frac{8\pi}{g^2} \int \! dt \,  \sqrt{(C/2)^2 - (\partial_0 \epsilon)^2} \, .
\label{monoeff}
\end{eqnarray}
This is an action of a relativistic particle, in other words, a 1-dimensional Nambu-Goto action. The speed of light is given by $C/2$. 

As a consistency check, let us calculate the Hamiltonian and compare it with the dyon mass. The Hamiltonian calculated from the moduli effective action
(\ref{monoeff}) is
\begin{eqnarray}
H = \frac{8\pi(C/2)^2}{g^2 \sqrt{(C/2)^2 - (\partial_0 \epsilon)^2}} \, .
\end{eqnarray}
Substituting the relation (\ref{idente}), this Hamiltonian is written
with $\gamma$ as
\begin{eqnarray}
H = \frac{4\pi C}{g^2} \cosh\gamma  
= \frac{4\pi C'}{g^2} \cosh^2\gamma \, 
\end{eqnarray}
which is exactly equal to the Julia-Zee dyon mass.
So, we conclude that the moduli effective action (\ref{monoeff})
describes correctly the dynamics of the moduli.

It is intriguing to note that the speed limit $C/2$ in the internal space
turns out to be equal to the mass of the W-bosons. For the case of 
the ${\mathbb C}P^1$ domain walls the speed limit is given by the
mass of the original scalar field, and we find a universal feature here :
The internal speed limit is given by the mass of the original 
massive field, per a second.

It might be interesting to consider the BPS 't Hooft-Polyakov monopole in the Higgs phase 
in ${\cal N}=2$ supersymmetric QCD with $N_{\rm F} = N_{\rm C} = 2$  \cite{Tong:2003pz}.  
There, the vacuum expectation value $C$ of the adjoint 
field in an $SU(2)$ vector multiplet is determined by a fundamental quark 
mass matrix $M = m \sigma_3$ as $\left<\phi\right> = M$.
Namely, we have $C = 2m$.
Since the monopole in the Higgs phase is pierced by
a squeezed magnetic flux, a vortex string, the moduli space is ${\mathbb R} \times S^1$.
Assuming the effective action (\ref{monoeff}) is valid even in the Higgs phase,
the speed limit in the internal $S^1$ space can be read as $\p_0 \epsilon = m$. In \cite{Tong:2003pz}, 
it was found that the monopole can be identified with a kink inside the vortex string. Namely, the kink is 
a topological soliton in $1+1$ dimensional massive ${\mathbb C}P^1$ model which is the low energy effective action
of the internal moduli of a non-Abelian vortex \cite{Hanany:2003hp,Auzzi:2003fs,Shifman:2004dr,Isozumi:2004vg}.
The moduli space of the kink in $1+1$ dimensions is ${\mathbb R}^1 \times S^1$ as we explained in the main body of the paper.
More concretely, the massive ${\mathbb C}P^1$ sigma model for the monopole is given by $F = \frac{4\pi}{g^2}(1+|\phi|^2)^{-2}$
in Eq.~(\ref{cp1}). This leads to the Nambu-Goto action (\ref{CP1s}) with the factor $m$ is multiplied by $4\pi/g^2$.
It gives the speed limit $m$ and it is 
identical to the Nambu-Goto action (\ref{monoeff}) with $C = 2m$.


\end{document}